\documentclass[fleqn,usenatbib]{mnras}

\usepackage{newtxtext,newtxmath}
\usepackage{orcidlink}

\usepackage[T1]{fontenc}

\DeclareRobustCommand{\VAN}[3]{#2}
\let\VANthebibliography\thebibliography
\def\thebibliography{\DeclareRobustCommand{\VAN}[3]{##3}\VANthebibliography}

\usepackage{graphicx}	
\usepackage{amsmath}	

\usepackage{booktabs}
\usepackage{threeparttable}
\graphicspath{{./}{figures/}}
\usepackage{soul}
\soulregister\cite7 
\soulregister\citep7 
\soulregister\citet7 
\soulregister\ref7 

\newcommand{\raa}{RAA}

\usepackage{array}

\title[Distribution of HI 21cm intervening absorbers]{Statistical distribution of HI 21cm intervening absorbers as potential cosmic acceleration probes}

\author[Chang-Zhi Lu et al.]{
	Chang-Zhi Lu$^{~\orcidlink{0000-0001-5834-6459}}$,$^{1,2}$
	Tingting Zhang,$^{3}$\thanks{E-mail: 101101964@seu.edu.cn(TTZ)}
	Tong-Jie Zhang$^{~\orcidlink{0000-0002-3363-9965}}$,$^{1,2,4}$\thanks{E-mail: tjzhang@bnu.edu.cn(TJZ)}
	\\
	$^{1}$Institute for Frontiers in Astronomy and Astrophysics, Beijing Normal University, Beijing 102206, China\\
	$^{2}$Department of Astronomy, Beijing Normal University, Beijing 100875, China\\
	$^{3}$College of Command and Control Engineering, PLA Army Engineering University, Nanjing 210017, China\\
	$^{4}$Institute for Astronomical Science, Dezhou University, Dezhou 253023, China\\
}

\date{Accepted 2023 March 10. Received 2023 March 10; in original form 2022 October 27}

\pubyear{2023}

\begin{document}
\renewcommand\arraystretch{1.2}
\label{firstpage}
\pagerange{\pageref{firstpage}--\pageref{lastpage}}
\maketitle

\begin{abstract}
	Damped Lyman-$\alpha$ Absorber (DLA), or HI 21cm Absorber (H21A), is an important probe to model-independently measure the acceleration of spectroscopic velocity ($v_\mathrm{S}$) via the Sandage-Loeb (SL) effect. Confined by the shortage of DLAs and Background Radio Sources (BRSs) with adequate information, the detectable amount of DLAs is ambiguous in the bulk of previous work. After differing the acceleration of scale factor ($\ddot{a}$) from the first order time derivative of spectroscopic velocity ($\dot{v}_\mathrm{S}$), we make a statistical investigation of the amount of potential DLAs in the most of this paper. Using Kernel Density Estimation (KDE) to depict general redshift distributions of BRSs, observed DLAs and a DLA detection rate with different limitations (1.4GHz flux, HI column density and spin temperature), we provide fitted multi-Gaussian expressions of the three components and their 1$\sigma$ regions by bootstrap, with a proportional constant of H21As in detected DLAs, leading to the measurable number predictions of H21As for FAST, ASKAP and SKA1-Mid in HI absorption blind survey. In our most optimistic condition ($F_\mathrm{1.4GHz}$>10mJy, $N_\mathrm{HI}>2\times10^{20}\mathrm{cm^{-2}}$ and $T_\mathrm{S}$>500K), the FAST, AKSAP and SKA1-Mid would probe about 80, 500 and 600 H21As respectively.
\end{abstract}




\section{Introduction}
An accelerating expansion is happening in our Universe, whose acceleration has not been measured model-independently up to now. An unprecedentedly precise observational determination will provide more clues to the fundamental modeling of the expanding mechanism, such as the most recognized dark energy or many candidates of Equation of State.

Common cosmological probes offer the model-dependent acceleration of scale factor($\ddot{a}$), while the first order time derivative of spectroscopic velocity ($\dot{v}_\mathrm{S}$) of objects faithfully tracing the Hubble flow in real time are still beyond our reach. Proposed by \cite{sanda62apj} and improved by \cite{loeb98apj}, the redshift drift, namely the Sandage-Loeb (SL) effect is a model-independent probe of $\dot{v}_\mathrm{S}$ and free of cosmic geometry, used to distinguish cosmological models\citep{codur21prd,mishr22prd} and explore the inhomogeneity and anisotropy of the Universe\citep{thoma22arx,heine22jcap}. \cite{mores22arx} recorded more details of redshift drift, and \cite{melia22ejph} discussed the significant differences between a zero and non-zero redshift drift universe.

Direct measurement of redshift drift mainly depends on lyman-$\alpha$ forest (optical method) and Damped Lyman-$\alpha$ Absorbers (DLAs, radio method, our main concern in this paper). DLAs contain abundant dense HI gas (column density $N_\mathrm{HI}>2\times10^{20}\mathrm{cm^{-2}}$) which absorbs Lyman-$\alpha$ photons (optical) and most 21cm radiation (radio) from radio sources in their local rest frame.

There are many studies about the physical properties of DLAs. \cite{zwaan05mn} analysed 355 HI 21cm line of nearby extragalaxies (z$\approx$0) from Westerbork Synthesis Radio Telescope, showing that their physical properties and DLAs incident rate are aligned with those of high-z DLAs (z>2) from quasar spectra. \cite{procha09apj} researched HI column density distribution function in comoving redshift f($N_\mathrm{HI}$,X) from a DLA survey in SDSS DR5, and argued that HI distribution barely evolves due to similar shapes of f($N_\mathrm{HI}$,X) in redshift from 2.2 to 5.5. Using statistical sub-sample from BOSS(a part of SDSS-III) quasar spectra at z>2, \cite{noter12aa} probed the $N_\mathrm{HI}$ distribution at $\langle z\rangle$=2.5, and found it matches the observed opacity-correct distribution well. \cite{braun12apj} used opacity-corrected high-resolution images of local extragalaxies to generate f($N_\mathrm{HI}$,X) at z$\approx$0. The distributions f and the number of equivalent DLAs show systematic decreases compared with their high-z values. By defining a measurable covering factor ($C_\mathrm{f}$), \cite{kanek14mn} carefully studied the DLA's spin temperature ($T_\mathrm{S}$) distribution in low and high redshift, within and beyond our galaxy, and the correlation between itself and $N_\mathrm{HI}$, $C_\mathrm{f}$ and metallicity [Z/H]. Using quasar spectra from the Hubble Space Telescope archive, \cite{neele16apj} conducted the first DLA blind survey at z<1.6 and analysed f($N_\mathrm{HI}$, X) in a specific range of $N_\mathrm{HI}$. Their DLA incidence was a bit lower but consistent with other results and uncertainties. \cite{rao17mn} used 70 MgII-preselected DLAs (with z<1.65) to describe DLA number density in 0<z<5 combing the z=0 modification and extra high-z DLAs. From SDSS-III DR12 DLAs data. \cite{bird17mn} revised f($N_\mathrm{HI}$,X) and dN/dX, especially the latter at z>4. Curran found that the reciprocal of $T_\mathrm{S}$ can trace the star formation density $\psi^*$ \citep{curran17aa}, and the fraction of the cold neutral medium ($\int \tau d \nu$/$N_\mathrm{HI}$) has relation to $\psi^*$ \citep{curran17mn}. \cite{grasha20mn} evaluated f($N_\mathrm{HI}$,X) per comoving redshift and per column density interval from detected absorptions and upper limits from non-detection regions. These works provided precious DLA samples and studied their multi-dimensional features, such as $N_\mathrm{HI}$, $T_\mathrm{S}$ and f($N_\mathrm{HI}$,X), which have been used in DLA amount predictions. Integrating both the linearly interpolated f($N_\mathrm{HI}$, $z$) and the simulated completeness function C($N_\mathrm{HI}$, $T_\mathrm{S}$, $v_\mathrm{FWHM}$, $z$) built from the FLASH early survey \citep{allison20mn}, they explored the posterior distribution of spin temperature \citep{allison21mn} considering different observations \citep{sadler20mn}, and the possible detection number \citep{allison22pasa} of intervening and associated HI 21cm absorbers given different constraints.

The first approach to $\dot{v}_\mathrm{S}$, Lyman-$\alpha$ forests (LFs), was carefully investigated by \cite{liske08mn} for the next generation instrument ESO-ELT and was recently renewed by \cite{dong22arx}. Optical Lyman-$\alpha$ absorbers are often discovered in intergalactic medium and likely have less peculiar acceleration \citep{cooke20mn}. With abundant Lyman-$\alpha$ absorption in line forests, the LFs approach gathers much attention like the Cosmic Accelerometer project \citep{eiken19baas}, the ACCELERATION programme \citep{cooke20mn}, ESPRESSO and NEID spectrographs \citep{chakr22arx}. Interestingly, \cite{estev21mn} concluded that measuring redshift drift and constraining cosmological parameters is a dilemma. However, confined by the earth's ionosphere, ground-based observations only receive LFs photons from $z\gtrsim1.65$ \citep{klo15aaska}, which corresponds to the decelerating expansion and jerk era.

The second approach, HI 21cm Absorbers (H21As), which are usually discovered in DLAs, was first attempted practically by \cite{darli12apj}, in which they gave the best constraint (until now) on redshift drift of three magnitudes larger than theoretical prediction. Long-term frequency stability was verified in GBT and their used H21As. Not all DLAs (H21As) are suitable for $\dot{v}_\mathrm{S}$ observations. A common classification \citep{curran16mn} divides DLAs into two types: associated (A-type, which are near the radio sources with statistically shallower and wider absorption profiles) and intervening (I-type, which are remote from the sources with deeper and narrower profiles), implying that I-type DLAs suffer less local effects and locate in colder and quieter regions, with fewer inner collisions and outside radiation. Therefore the I-type is more ideal for the observation of cosmic acceleration, and we abbreviate it as DLA hereafter. \cite{jiao20jcap} made an HI 21cm absorption spectral observation in PARKES, advocating the necessity of consecutive (decade) high-resolution spectral observations against the high-velocity uncertainty. \cite{lu21arx} made a high-accuracy HI 21cm spectral observation with FAST as a preliminary effort to obtain a snapshot of the S-L signal, introducing semi-theoretical velocity uncertainties in one epoch. Further observations for stricter constraints are still applied and prepared.

The redshift number density of potential DLAs were produced by \cite{yu14prl,yu17raa} and \cite{jiao20jcap}, where they focused on CHIME, Tianlai and FAST respectively. \cite{zhang21mn} estimated the number of HI absorption lines from the radio luminosity function of radio-load AGNs. And recently \cite{allison22pasa} evaluated it as mentioned before. The HI 21cm absorption surveys have found few new DLAs \citep{dutta17mn}, and the radio surveys obtained massive samples extended to deeper view and fainter sources \citep{matth21apj}. Therefore it necessitates checking as many datasets to renew the redshift number density of BRSs and DLAs. Meanwhile, many HI 21cm absorption survey programs are on schedule or undergoing, such as the First Large Absorption Survey in HI (FLASH) \citep{allison22pasa} in ASKAP telescope, MeerKAT Absorption Line Survey (MALS) \citep{gupta16mks,gupta21apj} in MeerKAT telescope, Widefield ASKAP L-band Legacy All-sky Blind surveY (WALLABY) \citep{kori20apss} in ASKAP too, will flesh our understanding of 21cm DLAs. Any advance in these aspects would modify the final anticipation of cosmic acceleration experiments in the radio approach.

In this paper, the difference in definition between the measured spectral acceleration ($\dot{v}_\mathrm{S}$) and the real cosmic (scale factor) acceleration ($\ddot{a}$) is stressed again in sec \ref{sec2}. With KDE to depict data and bootstrap to give the 1$\sigma$ errors, we study the redshift number density of Background Radio Sources (BRSs) in sec \ref{sec3.1} and observed DLAs in sec \ref{sec3.2} where we also explore DLA detection rate function and the fraction of H21As to DLAs, and estimate the detectable amount of potential H21As for FAST, ASKAP and SKA1-Mid in sec \ref{sec3.3}. Our discussion and conclusion are in sec \ref{sec4} and \ref{sec5}. All the calculation in this paper is based on a fiducial Planck18 $\Lambda$CDM model ($\Omega_\mathrm{K0}=\Omega_\mathrm{R0}=0$, $\Omega_\mathrm{M0}=0.315$, $H_0=67.4\mathrm{km\ s^{-1}\ Mpc^{-1}}$) \citep{planck20aa}.

\section{COSMIC ACCELERATION}\label{sec2}
For many papers containing the necessary formulae and derivations such as \cite{liske08mn}, we give a brief description with a standard $\Lambda$CDM model.

The expansion rate in $\Lambda$CDM is:
\begin{equation}\label{eq1}
	E(z)=\sqrt{\Omega_\mathrm{R0}(1+z)^4+\Omega_\mathrm{M0}(1+z)^3+\Omega_\mathrm{K0}(1+z)^2+\Omega_\mathrm{\Lambda0}},
\end{equation}
where $z$ is redshift, and $\Omega_\mathrm{R0},\Omega_\mathrm{M0},\Omega_\mathrm{K0},\Omega_\mathrm{\Lambda0}$ are today's density parameters of radiation, matter, curvature and dark energy respectively. The redshift drift of Hubble-flow tracers is:
\begin{equation}\label{eq2}
	\Delta z\approx\frac{\dot{a}_0-\dot{a}(z)}{a(z)}\Delta t_0=H_0[1+z-E(z)]\Delta t_0,
\end{equation}
where $t_0=t_{\mathrm{obs}}$ is the time of observer, $a_0=a(t_0)$ is today scale factor, the dot in $a$ represents the derivative with respect to $t_0$ , and $H_0$ is Hubble constant.

The change of its spectroscopic radial velocity ($v_\mathrm{S}=cz$) is:
\begin{equation}\label{eq3}
	\Delta v_{\rm S}\approx\frac{c}{1+z}\Delta z\approx c[\dot{a}_0-\dot{a}(z)]\Delta t_0=cH_0[1-\frac{E(z)}{1+z}]\Delta t_0.
\end{equation}
The velocity drift is an indicator of $\dot{a}$ differences ($\dot{a}_0-\dot{a}(z)$). And according to eq. \ref{eq3}, we can use measured $\dot{v}_{\rm S}$ to further infer $\Delta\dot{a}$ and $\ddot{a}$ model-independently.

When we explain the cosmic expansion with general relativity (GR), the recession velocity is \citep{davis01aipc}:
\begin{equation}\label{eq4}
	v_{\rm G}=\dot{a}(z)D_{\rm C}(z)=\frac{c}{H_0}\dot{a}(z)\int_0^z\frac{dz'}{E(z')},
\end{equation}
where $D_{\rm C}(z)=c\int_0^z\frac{dz'}{H(z')}$ is the comoving distance. The first order time derivative of $v_{\rm G}$ is \citep{jiao20jcap}:
\begin{equation}\label{eq5}
	\dot{v}_{\rm G}=\ddot{a}(z)D_{\rm C}(z)+\dot{a}(z)\dot{D}_{\rm c}(z)=\ddot{a}(z)D_{\rm C}(z)+\dot{a}(z)\frac{dD_{\rm c}(z)}{dz}\frac{dz}{dt_0}.
\end{equation}
In order to conveniently plot $\dot{v}_{\rm G}$, from the expression of deceleration parameter $q(z)$, we could write down $\ddot{a}(z)$:
\begin{equation}\label{eq6}
	q(z)=\frac{1+z}{2E^2(z)}\frac{dE^2(z)}{dz}-1,
\end{equation}
\begin{equation}\label{eq7}
	\ddot{a}(z)=-H_0^2q(z)a(z)E^2(z).
\end{equation}

Only $\ddot{a}$ can depict the physical process of deceleration or acceleration of the universe expansion, and have a consistent value at any point from any distance in the same moment, for the scale factor is generalized to the whole universe.

The observed $\dot{z}$ (or $\dot{v}_\mathrm{S}$) involves a comparison within a pair of space-time spots (today-observer and past-source). For a fixed source, its $\dot{z}$ (or $\dot{v}_\mathrm{S}$) would change with the observer's position selection. Thus it is just a relative measurement of the apparent velocity changes for a specific local observer. Although we do measure it, it does not represent the real acceleration of the Universe's expansion. Moreover, through a secular (decade) observation, we can derive a time-averaged $\dot{a}$ drift ($\overline{\Delta\dot{a}}$) from $\Delta v_\mathrm{S}$ in eq. \ref{eq3}.

\begin{figure}
	\centering
	\includegraphics[scale=0.6]{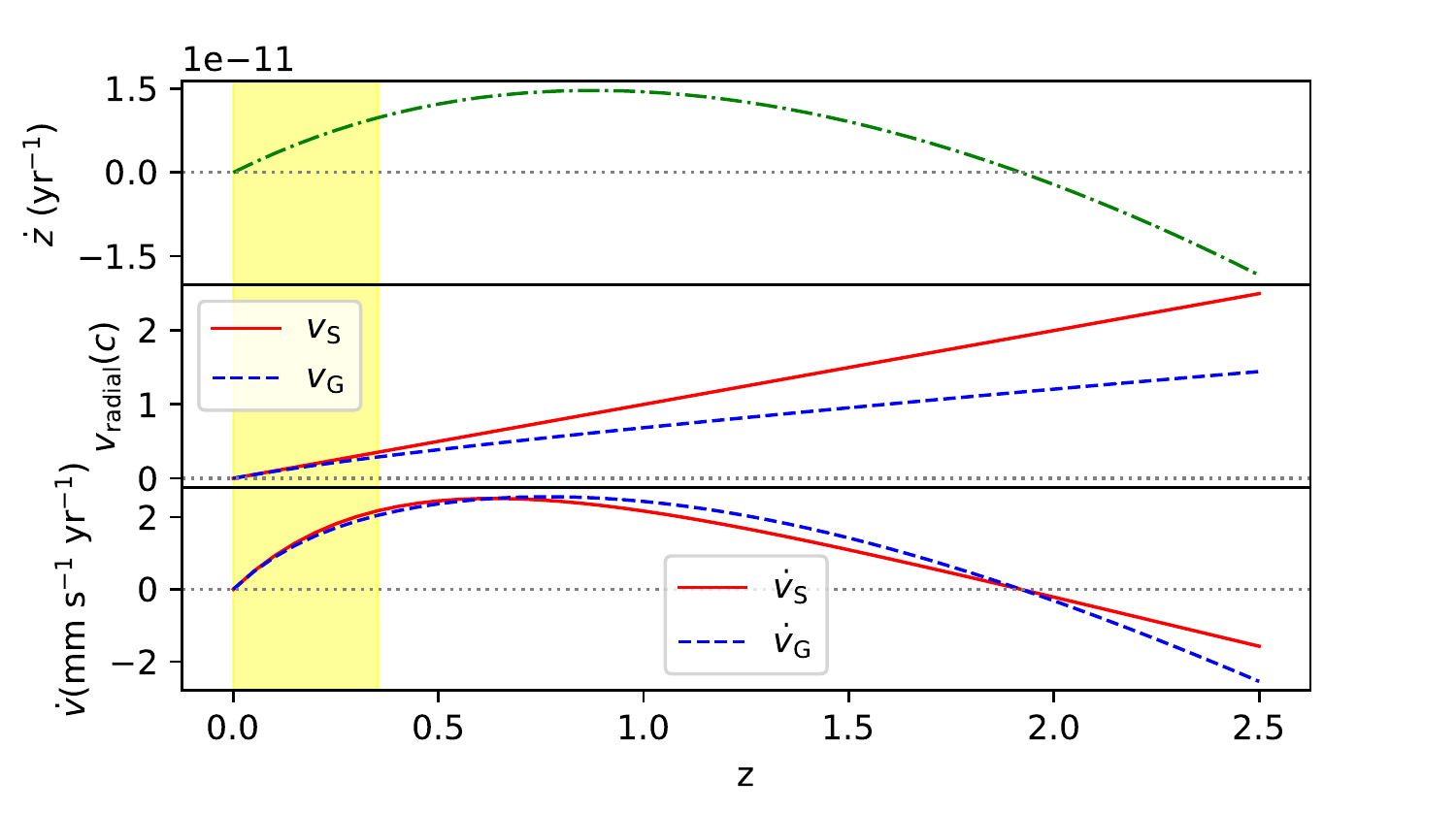}
	\caption{Several physical quantities versus redshift. In the upper panel, the green dashed-dotted line is the redshift drift (eq. \ref{eq2}). In the middle panel, the red solid and blue dashed lines are the spectroscopic radial and GR recession velocities (eq. \ref{eq4}). In the bottom panel, we show their corresponding velocity drifts (eq. \ref{eq3} and \ref{eq5}). The pale yellow block across three panels covers the redshift range of FAST ($0\sim$0.352) \citep{li18imm}.}\label{fig1}
\end{figure}

From Figure \ref{fig1}, the zero-points of $\ddot{a}$ and $\dot{z}$ (or low-redshift approximated $\dot{v}_\mathrm{S}$) are very different, where the former ($z_\mathrm{a0}\approx0.6$) relies on $q(z)$ and the latter ($z_\mathrm{z0}\approx1.9$) depends on $1+z-E(z)$.

\begin{figure}
	\centering
	\includegraphics[scale=0.6]{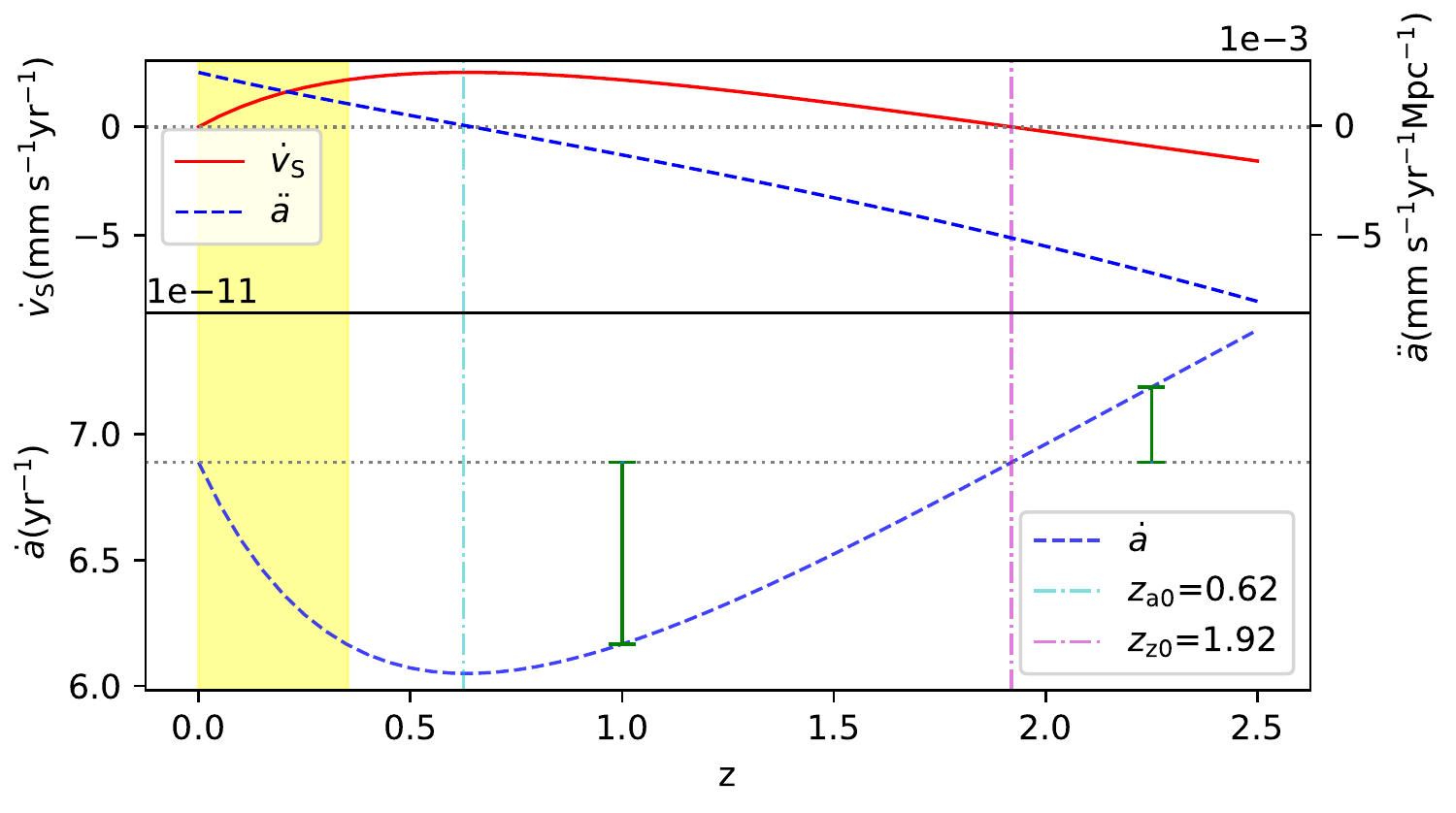}
	\caption{The measurable $\dot{v}_\mathrm{S}$ and theoretical $\ddot{a}$. The upper panel contains $\ddot{a}$ (the blue dashed line, eq. \ref{eq7}) and $\dot{v}_\mathrm{S}$ (the red line, eq. \ref{eq3}). In the lower panel, the blue dashed line is $\dot{a}=a(z)H(z)$. The left vertical cyan dashed-dotted line is the zero-point of $\ddot{a}$, while the right vertical magenta dashed-dotted line is the zero-point of $\dot{z}$ (or $\dot{v}_\mathrm{S}$). We use two short green bars to emphasise the positive or negative value of $\dot{v}_\mathrm{S}\propto(\dot{a}_\mathrm{0}-\dot{a}_\mathrm{z})$. The pale yellow block is the FAST redshift coverage.}\label{fig2}
\end{figure}

From the upper panel in Figure \ref{fig2}, the universe expands in acceleration at $z<z_\mathrm{a0}$ ($\ddot{a}$>0), and the universe expands in deceleration at $z>z_\mathrm{a0}$ ($\ddot{a}$<0). But $z_\mathrm{z0}$ cannot distinguish between the two states. For example, when $z<z_\mathrm{z0}$, the observed $\dot{v}_\mathrm{S}\propto(\dot{a}_\mathrm{0}-\dot{a}_\mathrm{z})$ is always positive, but it does not mean the whole universe expanding acceleratingly from $z_\mathrm{z0}$ to $z=0$. Instead, it only proves that the observed point is escaping us visually during this era.

From the lowest panel in Figure \ref{fig2}, if our Universe is genuinely dominated by a $\Lambda$CDM model, we could safely research spectroscopic acceleration ($\dot{v}$) with good approximation at $z\lesssim2$.

After distinguishing $\dot{v}_{\rm S}$ and $\ddot{a}$, in the following sections, we use existing radio sources and H21As samples to make a realistic number prediction, for the purpose that one can observe them to better constrain $\dot{v}_{\rm S}$ via the radio approach.

\section{EXPECTATION OF HI 21cm absorption systems}\label{sec3}
The redshift number density of potential H21As in every square degree of the sky with three lowest detection limitations, i.e. flux density ($F$) of BRSs, HI column density ($N_\mathrm{HI}$) and spin temperature ($T_\mathrm{S}$) of DLAs, could be expressed as
\begin{equation}\label{eq8}
\begin{aligned}
	n_\mathrm{21S}(z,F,N_\mathrm{HI},T_\mathrm{S})&=
	\eta_\mathrm{21}(z)\int_{N_\mathrm{HI}}^\infty\int_{T_\mathrm{S}}^\infty\kappa(z,N',T')n_\mathrm{D}(z,N')dN'dT'\\ &\times\int_{F}^\infty\int_z^\infty n_\mathrm{R}[z'+\Delta Z(z'),F']dz'dF'.
\end{aligned}
\end{equation}

$n_\mathrm{R}(z,F)$ is the number density of BRS per square degree related with redshift $z$ and observed 1.4GHz flux density $F$. $n_\mathrm{D}(z,N_\mathrm{HI})$ is the number density of observed DLAs in every possible sightline directing toward a BRS with $z$ and HI column density ${N_\mathrm{HI}}$. $\kappa(z,N_\mathrm{HI},T_\mathrm{S})$ is the DLA detection rate at a given ${N_\mathrm{HI}}$ and $T_\mathrm{S}$ level. $\eta(z)$ is the proportion of H21As in DLAs with $z$. Moreover, $\Delta Z(z)$, used in \cite{allison21mn}, excludes the background galaxy near its foreground DLA within the radial velocity of 3000$\mathrm{km/s}$:
\begin{equation}\label{eq9a}
	\Delta Z(z)=3000(1+z)/c,
\end{equation}
where $c$ is the speed of light.

\subsection{Prediction of BRSs}\label{sec3.1}
The common redshift distribution of radio sources in predicting potential-DLA detectable number is \citep{dezot10aapr}:
\begin{equation}\label{eq9b}
	n_\mathrm{R}(z)=1.29+32.37z-32.89z^2+11.13z^3-1.25z^4,
\end{equation}
which came from a $z$-binned polynomial fitting of the CENSORS data \citep{brook08mn}. However, the polynomial will be negative when $z$ exceeds 3.5. Besides, the expression contains radio sources with $F_\mathrm{1.4GHz}\leq10\mathrm{mJy}$, which are faint and time-expensive in a blind survey to acquire prominent absorption lines.

The Combined EIS-NVSS Survey of Radio Sources (CENSORS) targeted the ESO Imaging Survey (EIS) Patch D, covering a $3\times2\mathrm{deg^2}$ field of view and 150 sources preselected from NRAO VLA Sky Survey (NVSS) \citep{condon98aj}. \cite{rigby11mn} presented a 135-sample subset of CENSORS at a $F_\mathrm{1.4GHz}$ completeness of 7.2mJy. The extra $F_\mathrm{1.4GHz}$ can provide more information and limitation in the BRS distribution.

Besides, \cite{marcab13arx} made a $z$-binned gamma fitting for CENSORS. When comparing the galaxy angular power spectrum via the Bayesian evidence test, the gamma fitting outperformed the polynomial one.

\subsubsection{More Datasets}\label{sec3.1.1}
The CENSORS provides many samples for radio-source redshift revolution. Nonetheless, can the $n_\mathrm{R}(z)$ from a $6\mathrm{deg^2}$ survey represent the whole sky situation? Considering the varied physical conditions toward every sky direction in every matter cluster, it still deserves verification with more radio datasets. The LBDS-Hercules \citep{waddi01mn} and CoNFIG-4 \citep{gendre10mn} covering the different sky areas are satisfied.

The Leiden-Berkeley Deep Survey (LBDS) Hercules has 64 radio sources with $F_\mathrm{1.4GHz}>2\mathrm{mJy}$ in a $2\mathrm{deg^2}$ sky, and we use the data collected by \cite{rigby11mn}. The Combined NVSS-FIRST Galaxies-4 (CoNFIG-4) contains 184 radio sources with $F_\mathrm{1.4GHz}>50\mathrm{mJy}$ in a $52\mathrm{deg^2}$ sky. The additional two sets greatly improve our analysis.

Firstly our radio datasets were made ten years ago, so we update them following the NVSS catalogue. New sources belonging to the sky coverage of CoNFIG-4 are listed in Table \ref{tab1}.

\begin{figure}
	\centering
	\includegraphics[scale=0.58]{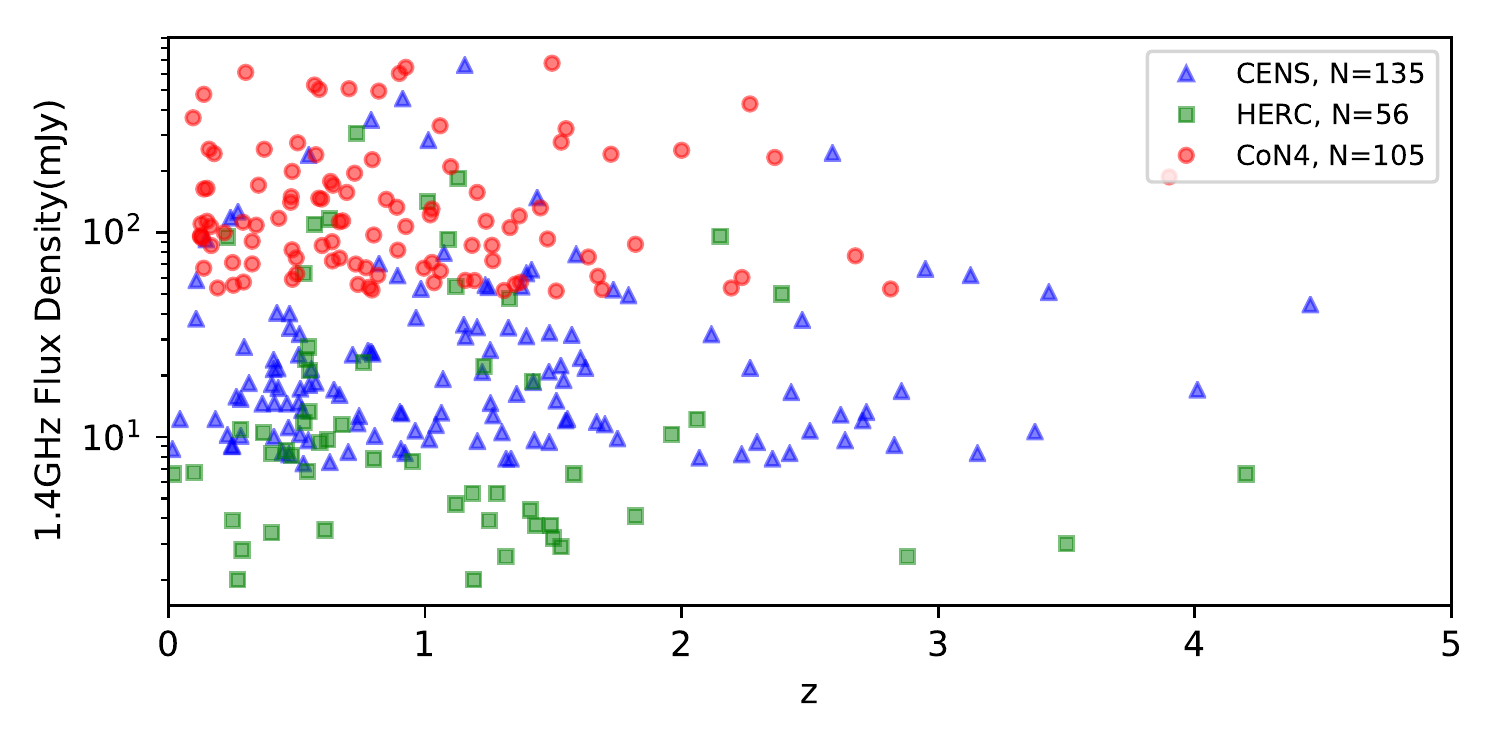}
	\caption{The distribution of three radio datasets. The blue triangles, green squares and red circles from CENSORS, Hercules and CoNFIG-4 respectively, are our used sources. Their amounts are listed in the legend.}\label{fig3}
\end{figure}

Then we plot the distributions of all BRSs with $z\geq0.1$ (beyond the local supercluster, but including a sample of z=0.0963 from CoNFIG-4) in Figure \ref{fig3}, and list the count of different 1.4GHz flux levels in Table \ref{tab2}. The flux counts are quite imbalanced between the CENSORS and Hercules, indicating a possible sky coverage bias.

To keep samples rich enough, we use CENSORS and Hercules to estimate the BRS $z$ probability density with $F_\mathrm{1.4GHz}\geq10\mathrm{mJy}$ and all sets to the $z$ density with $F_\mathrm{1.4GHz}\geq50\mathrm{mJy}$.

Although our CoNFIG-4 dataset contains the original 184 and extra 3 extra radio sources, only 105 of them have $z\geq0.0963$. Therefore we re-scale the sky coverage as $105/187*52\approx29.2\mathrm{deg^2}$, to counteract the lost $z$ information.
\begin{table*}
	\centering
	\caption{Additional radio sources in the sky coverage of CoNFIG-4}\label{tab1}
	\begin{threeparttable}
	\setlength{\tabcolsep}{5mm}{
		\begin{tabular}{ccccc}
			\toprule
			Source Name & RA Dec(J2000) & $z$ & $F_\mathrm{1.4GHz}$ & Reference$^1$\\
			& (h:m:s d:m:s) & & (mJy) & \\ 
			\midrule
			QSO J1407-0049 & 14 07 10.59 -00 49 15.3 & 1.51117 & 51.8 & [A09]\\
			SDSS J143031.30-000907.5$^2$ & 14 30 31.45 -00 09 08.0 & 2.8143 & 53 & [A15]\\
			NVSS J143403+010351$^2$ & 14 34 03.22 +01 03 51.5 & 1.06 & 64.9 & [T19]\\
			TXS 1423+019$^2$ & 14 26 30.42 +01 42 36.1 & 0.3263 & 90.8 & [G19]\\
			TXS 1408+016$^2$ & 14 11 08.29 +01 24 41.1 & 3.9 & 187.5 & [M19]\\
			TXS 1420+018$^2$ & 14 23 03.43 +01 39 58.7 & 1.1 & 210.4 & [M19]\\
			LBQS 1438+0210 & 14 40 59.50 +01 57 43.9 & 0.7944 & 227.6 & [A15]\\
			NVSS J140639-032430 & 14 40 59.50 +01 57 43.9 & 2 & 253 & [X14]\\
			TXS 1434-003$^2$ & 14 37 21.09 -00 33 18.1 & 0.0963 & 365.0 & [M19]\\
			TXS 1406+015$^2$ & 14 08 33.31 +01 16 22.1 & 0.9 & 601.3 & [S06]\\
			\bottomrule
		\end{tabular}}
		\begin{tablenotes}
			\footnotesize
			\item $^1$ [A09]\citep{abaza09apjs}, [A15]\citep{alam15apjs}, [T19]\citep{toba19apjs},\newline[G19]\citep{garon19aj}, [M19]\citep{ma19apjs}, [X14]\citep{xu14mn}, [S06]\citep{schmi06apj}.
			\item{$^2$} These radio sources were originally contained in CoNFIG-4 dataset but without $z$ information.
		\end{tablenotes}
	\end{threeparttable}
\end{table*}
\begin{table}
	\centering
	\caption{1.4GHz flux counts of datasets with $z\geq0.1$}\label{tab2}
	\begin{threeparttable}
		\begin{tabular}{cccc}
			\toprule
			dataset & N($F_\mathrm{1.4GHz}\geq50\mathrm{mJy}$) & N($F_\mathrm{1.4GHz}\geq10\mathrm{mJy}$) & N(all)\\
			\midrule
			CENSORS & 25 & 108 & 135 \\
			Hercules & 11 & 25 & 56 \\
			CoNFIG-4 & 105 & 105 & 105$^1$ \\
			\bottomrule
		\end{tabular}
		\begin{tablenotes}
			\footnotesize
			\item{$^1$} CoNFIG-4 data include a sample of z=0.0963.
		\end{tablenotes}
	\end{threeparttable}
\end{table}

\subsubsection{New Descriptions and One Example}\label{sec3.1.2}
Polynomial fitting often has intractable zero points. Specific fitting models such as gamma require more prior knowledge. And different-length $z$-bin would lose diverse information in the process.

Consequently, we introduce the Kernel Density Estimation (KDE) to implement a non-parametric fitting of $z$ probability density without bins. Particularly when using a Gaussian kernel, the final distribution can be extracted by multi-Gaussian fitting. In the next paragraphs, we will show the data process of CoNFIG-4 with $F_\mathrm{1.4GHz}\geq50\mathrm{mJy}$ as an example. For other datasets following the same procedure, we would give results directly.

The scale of abscissa would affect the profile's height, so we fix the gap of z as 0.1 at the start.

We use Gaussian kernel KDE (\textit{KernelDensity} in sklearn) to fit the CoNFIG-4 data (both $z$ and $F_\mathrm{1.4GHz}$). KDE has a key parameter 'bandwidth' equivalently controlling the width of a single Gaussian count unit, which replaces a single rectangle count unit in a histogram. Narrower bandwidth remains more features, but leads to a more flexible model and less generalization capability.

\begin{figure}
	\centering
	\includegraphics[scale=0.58]{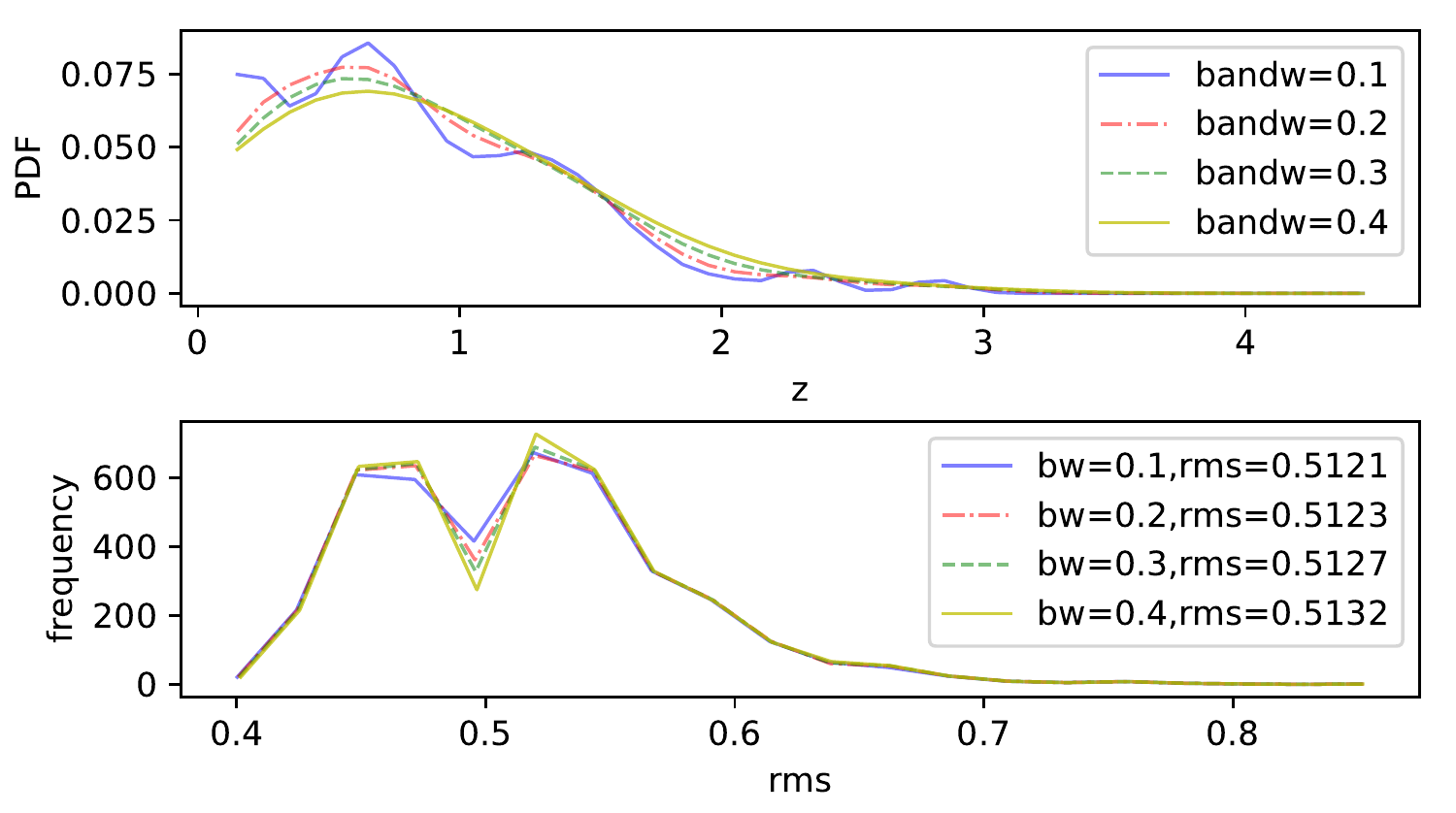}
	\caption{The different bandwidths in CoNFIG-4 KDE. In the upper panel, the blue, red dashed-dotted, green dashed and yellow solid curves are the PDF with bandwidths from 0.1 to 0.4. In the lower panel, four curves are the rms error distribution in 500 rounds' cross-validation, their colors correspond to the upper, and their mean rms are shown in the legend.}
	\label{fig4}
\end{figure}

For instance, we plot the fitted Probability Density Function (PDF) with several bandwidths in the upper panel of Figure \ref{fig4} and show their rms error count from 500 rounds' 8-fold random-shuffle cross-validation in its lower panel. The mean rms error decreases when the bandwidth diminishes, and the rms error distribution keeps stable. However, the 0.1-bandwidth at least needs a 5-Gaussian profile to fit. Thus we accept a bandwidth$\geq$0.2 but as small as possible and always prefer a simpler multi-Gaussian fitting.

Next, we explore what statistical model covers CoNFIG-4 better. Noticing the denser low-redshift region, we use exponential, gamma and Weibull distribution to fit the data.

\begin{figure}
	\centering
	\includegraphics[scale=0.58]{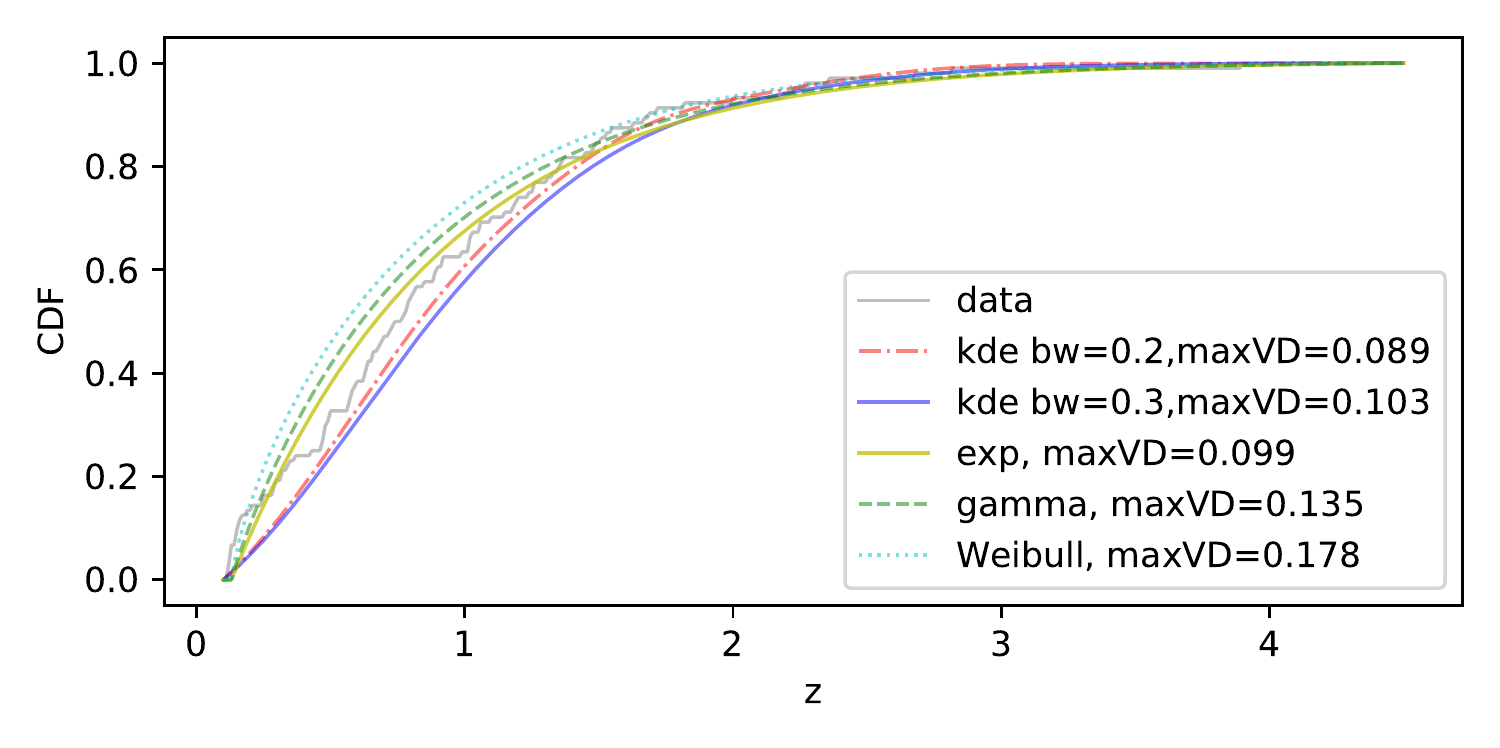}
	\caption{The CDF for manual K-S test. The gray, red dashed-dotted, blue, yellow, green dashed and cyan dotted curves are the CDFs of real data(CoNFIG-4), 2 KDEs(bw=0.2, 0.3), exp, gamma and Weibull respectively. Their maxVDs are listed in the legend.}
	\label{fig5}
\end{figure}

We use Kolmogorov-Smirnov(K-S) test (\textit{kstest} in scipy) to compare their fittings. If it returns a p-value higher than our expected level of significance ($\alpha=0.05$), we can regard the two arrays from the same distribution. It shows the p-values 0.0815 of the norm and 0.0573 of exponential (exp), while gamma and Weibull both have p-values lower than 0.05. But the auto K-S test can not process the KDE curve. Therefore in manual coding, we plot the Cumulative Distribution Function (CDF) of real 0.1-z-bin data, exponential, KDE curve of bandwidth 0.2 and 0.3 in Figure \ref{fig5}, and compare the maximum Vertical Distances (maxVDs) between real data CDF and tested CDF. Now we define a constant:
\begin{equation}\label{eq10}
	C(\alpha,m,n)=\sqrt{-\frac{1}{2}\ln(\frac{\alpha}{2})}\sqrt{\frac{m+n}{m*n}},
\end{equation}
where $\alpha$ is the significant level, $m$ and $n$ are the numbers of real samples and tested samples. We gain the KDE profile from real samples, so C(0.05,104,104)$\approx$0.1883. If maxVD is less than the constant C, we can treat the two arrays from the same distribution.

\begin{figure}
	\centering
	\includegraphics[scale=0.58]{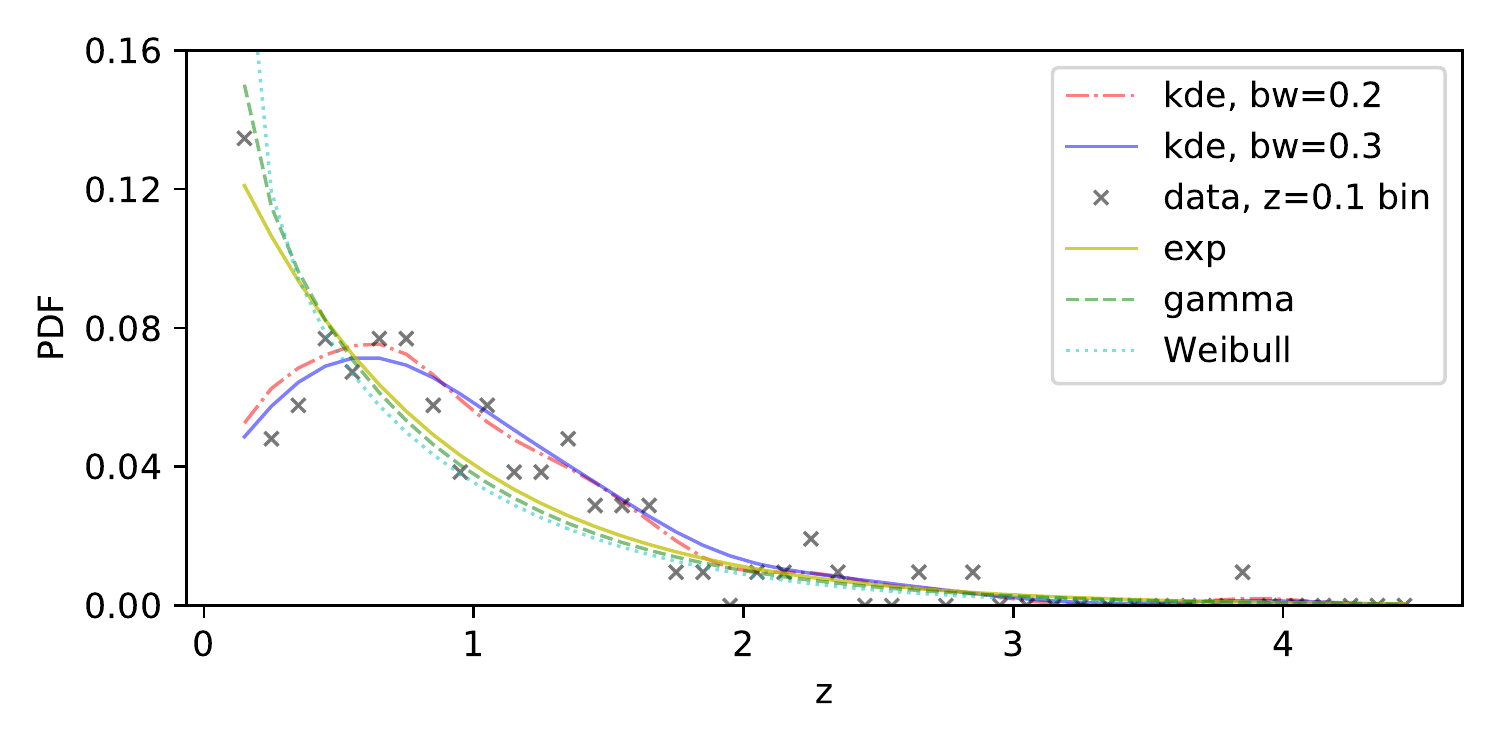}
	\caption{The fitted PDF of CoNFIG-4. The gray cross is the Probability Density Law(PDL) of the real data. The red dashed-dotted, blue, yellow, green dashed and cyan dotted curves are the PDFs of 2 KDEs (bw=0.2, 0.3), exp, gamma and Weibull separately.}
	\label{fig6}
\end{figure}

From Figure \ref{fig5}, the exp and 2 KDEs comparatively fit real data from the aspect of maxVD. We plot their PDFs in Figure \ref{fig6}, where the exp and 2 KDEs are distinct. The gamma and Weibull are worse than the exp because their p-values are lower than 0.05. Though the exp matches the peak point well, it behaves poorly at the adjacent redshifts, causing the overestimation in our most concerned low-redshift (z<1) region.

\begin{figure}
	\centering
	\includegraphics[scale=0.58]{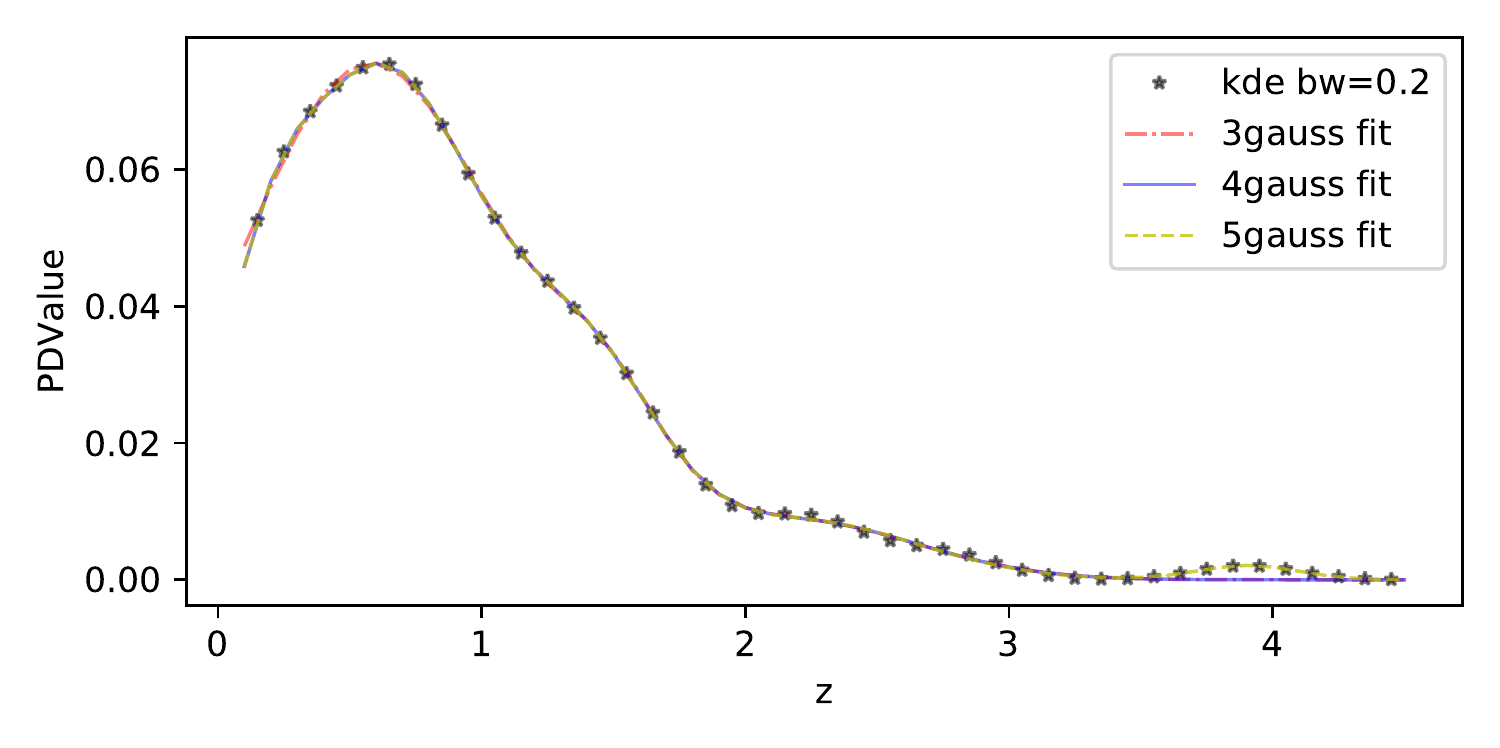}
	\caption{The multi-Gaussian fitting of KDE with bandwidth 0.2 for CoNFIG-4. The gray star is KDE data. The red dashed-dotted, blue and yellow dashed curves are 3-, 4- and 5-Gaussian models separately.}
	\label{fig7}
\end{figure}

Finally, multi-Gaussian fitting exports an analytic expression for the BRS integration in eq. \ref{eq8}. To balance the informative feature and simple model, we plot the 0.2-bandwidth fitting in Figure \ref{fig7} and choose the 3-Gaussian result.

Because we mainly concern with the KDE performance, we will ignore other statistical model fittings hereafter.

\subsubsection{Probability Density of BRSs}\label{sec3.1.3}
According to the H21A spectroscopic observation \citep{gereb15aa}, where 3 BRSs in all 32 ones showed $F_\mathrm{1.4GHz}\leq50$mJy (but $\geq35$mJy). Thus 50mJy can be a proper lower limit for the present observation capability. And we regard 10mJy as an optimistic capability for the next generation of radio telescopes or arrays.

\begin{figure}
	\centering
	\includegraphics[scale=0.58]{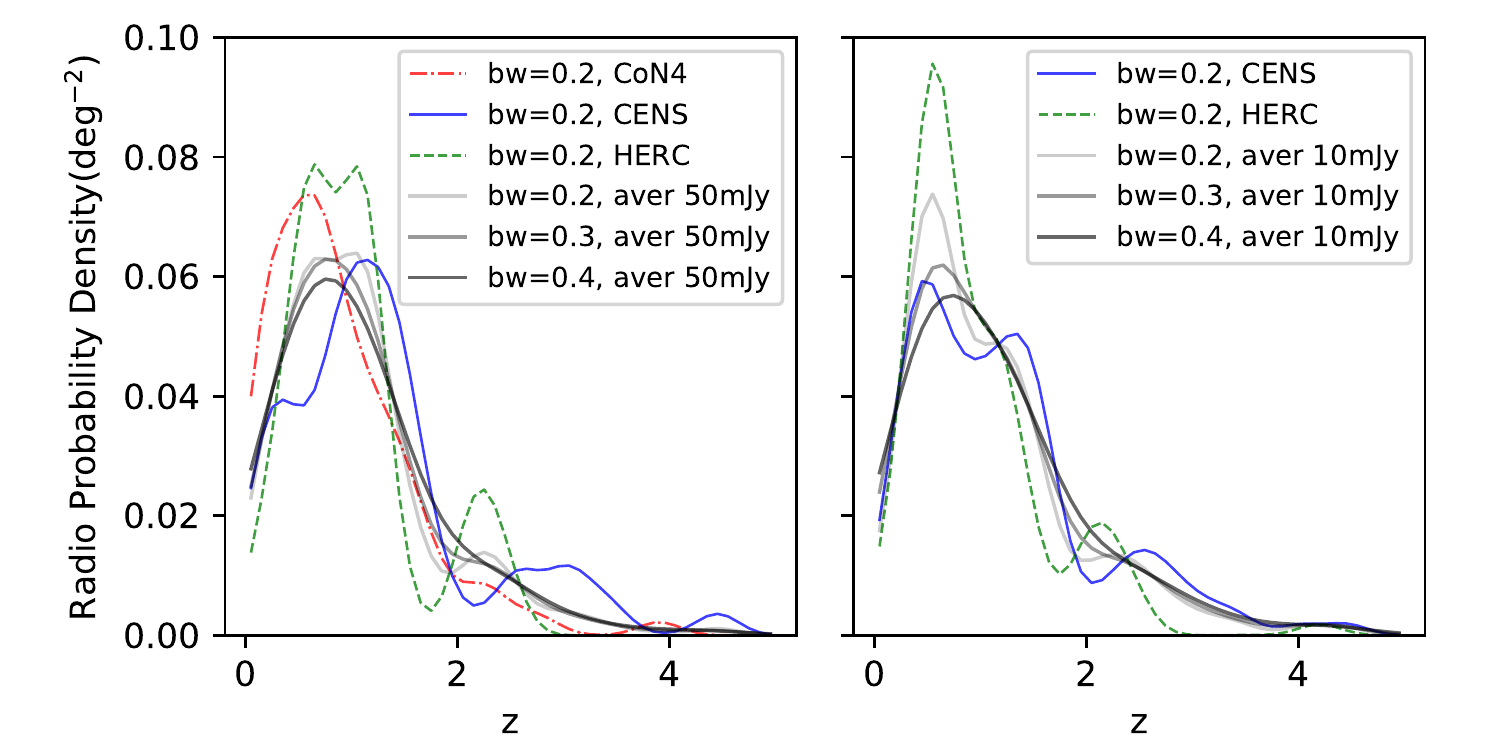}
	\caption{The multi-Gaussian fitting of the KDE curve with bandwidth 0.2 for all radio datasets. The left panel is for $F_\mathrm{1.4GHz}\geq50\mathrm{mJy}$ and the right is for $F_\mathrm{1.4GHz}\geq10\mathrm{mJy}$. The red dashed-dotted, blue and green dashed curves are the 0.2-bandwidth KDE results of CoNFIG-4, CENSORS and Hercules separately. The gray lines are the number-weighted KDEs with different-bandwidth components.}
	\label{fig7a}
\end{figure}

We make KDE for different radio sets, accumulate with two flux levels (50mJy and 10mJy), check with manual K-S test, average them with the weights of their number density per square degree sky ($\sigma_\mathrm{CO_{50}}$=3.5962, $\sigma_\mathrm{CE_{50}}$=4.1667, $\sigma_\mathrm{HE_{50}}$=5.5,$\sigma_\mathrm{CE_{10}}$=18, $\sigma_\mathrm{HE_{10}}$=12.5), and plot these in Figure \ref{fig7a}, where larger bandwidths relate to simpler models, but lose more features, especially merging the first two peaks, which heavily impact our estimation in the low-$z$ region. Therefore, we fix the bandwidth as 0.2.

\begin{figure}
	\centering
	\includegraphics[scale=0.58]{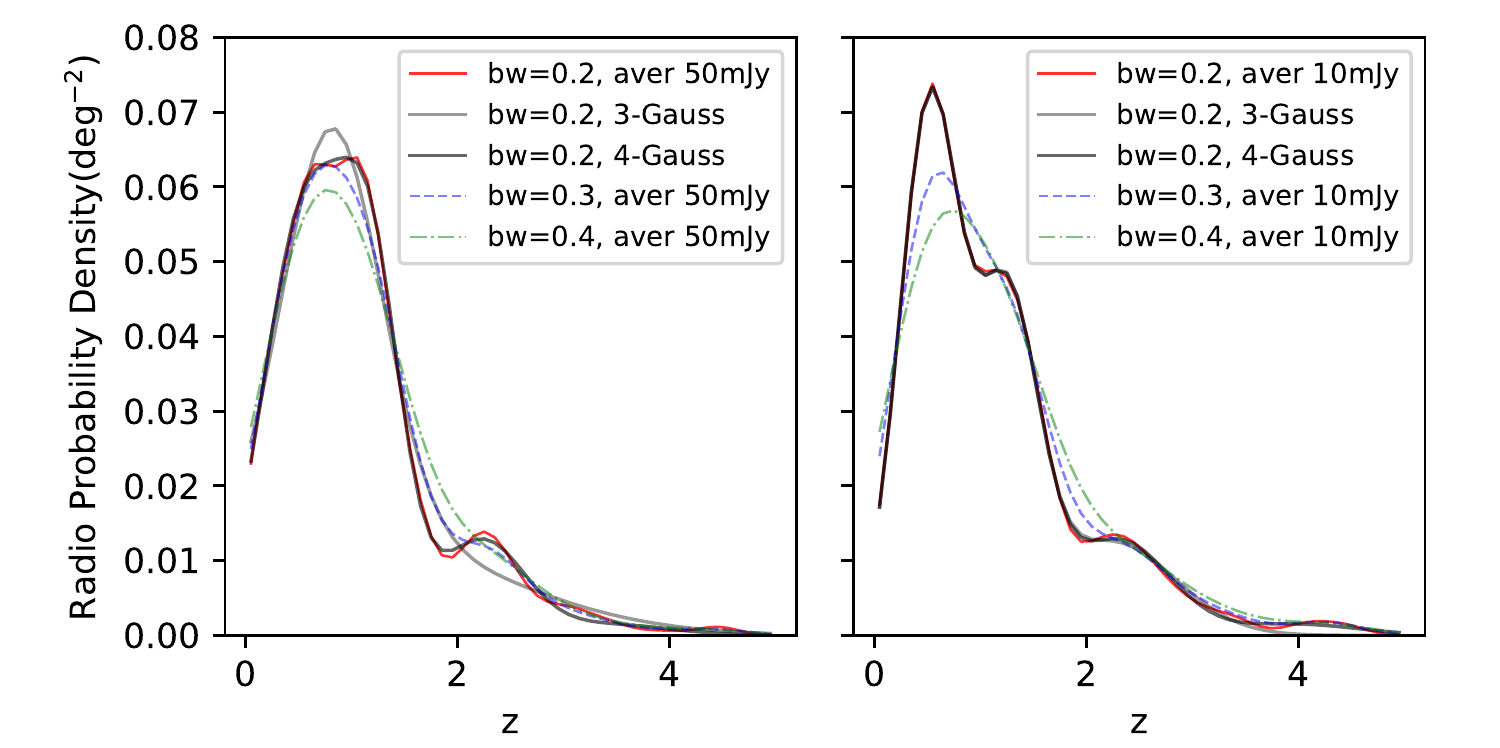}
	\caption{The multi-Gaussian fitting of averaged 0.2-bandwidth radio KDE curve. The left panel is for $F_\mathrm{1.4GHz}\geq50\mathrm{mJy}$ and the right is for $F_\mathrm{1.4GHz}\geq10\mathrm{mJy}$. The red, gray and black curves are the KDE averaged result and its KDE (bandwidth=0.2, 0.3) individually. The blue dashed and green dashed-dotted lines are the averaged other-bandwidth radio KDE.}
	\label{fig7b}
\end{figure}

We make multi-Gaussian fitting for a 2-dimension ($z$-$F_\mathrm{1.4GHz}$) 0.2-bandwidth KDE curves in Figure \ref{fig7b}. The 3-Gaussian is our prioritized option for simplicity. But in the high-$z$ region (z>4) of the 10mJy lower limit, it loses the last peak and causes an irreparable decrease compared with the 50mJy lower limit, even considering the 2 area number densities. Such inconsistency obliges us to add an extra Gaussian component to solve it, and the final 4-Gaussian fitting parameters and their 1$\sigma$ errors are given in Table \ref{tab3}.
\begin{table}
	\centering
	\caption{Multi-Gaussian fitting of averaged 0.2-bandwidth radio KDE curves}\label{tab3}
	\begin{threeparttable}
		\begin{tabular}{cccc}
			\toprule
			component & $a^1$ & $\mu^1$ & $\sigma^1$\\
			\midrule
			50mJy-1 & 0.0579$\pm$0.0011 & 0.5790$\pm$0.0187 & 0.3930$\pm$0.0123\\
			50mJy-2 & 0.0395$\pm$0.0030 & 1.1939$\pm$0.0111 & 0.2711$\pm$0.0084\\
			50mJy-3 & 0.0124$\pm$0.0002 & 2.2123$\pm$0.0168 & 0.4869$\pm$0.0208\\
			50mJy-4 & 0.0012$\pm$0.0002 & 4.0000$\pm$0.1029 & 0.5000$\pm$0.1178\\
			10mJy-1 & 0.0706$\pm$0.0003 & 0.5193$\pm$0.0023 & 0.2788$\pm$0.0016\\
			10mJy-2 & 0.0447$\pm$0.0003 & 1.2491$\pm$0.0033 & 0.3031$\pm$0.0042\\
			10mJy-3 & 0.0127$\pm$0.0001 & 2.2949$\pm$0.0122 & 0.5049$\pm$0.0147\\
			10mJy-4 & 0.0015$\pm$0.0001 & 4.0488$\pm$0.0515 & 0.5000$\pm$0.0588\\
			\bottomrule
		\end{tabular}
		\begin{tablenotes}
			\footnotesize
			\item{$^1$} Gaussian function is $f(z)=\frac{a}{\sqrt{2\pi}\sigma}\exp{[\frac{(z-\mu)^2}{2\sigma^2}]}$.
		\end{tablenotes}
	\end{threeparttable}
\end{table}

The fourth Gaussian in 2 KDEs are constrained insufficiently: the $a$, $\mu$ and $\sigma$ approach their set borders of 0.15 (a+), 4($\mu$-) and 0.5 ($\sigma$+). If we loosen the borders, the 4-Gaussian tends to become the 3-Gaussian again. Actually, our intention is to derive an approximate KDE expression for numerical integration, and the 4-Gaussian fitting traces the KDE better than the 3-Gaussian, which is enough. Besides, we find that the former 3 components (8 parameters, excluding $\sigma_\mathrm{3}$) are insensitive to the slight change of limitation (keeping $\mu_\mathrm{4}$>3.5), varying less than 5\% ($\sigma_\mathrm{3}$ is 20\%) and 10\% in the fitted values and errors. Moreover, the curve tails (z>4) are quite small, only covering less than 1\% of the total area under the curve. Therefore we choose to use the 4-Gaussian.

The errors in Table \ref{tab3} come from our used samples, and do not reflect the uncertainty including sampling, which need the bootstrap method to convey. The bootstrap randomly re-samples from the original dataset ($A_\mathrm{0}$) with replacement forming a number of new datasets ($\mathscr{A}_\mathrm{N}=\{A_\mathrm{i}\}_\mathrm{N}$, often N$\geq$1000) with the same data volume. One conducts statistical inference from the newly generated sets ($\mathscr{A}_\mathrm{N}$) assuming the $A_\mathrm{0}$ and $\mathscr{A}_\mathrm{N}$ from the almost same underlying population. Thus, it provides a more robust 1$\sigma$ estimation from the potential sample variations, particularly for small sample data.

\begin{figure}
	\centering
	\includegraphics[scale=0.58]{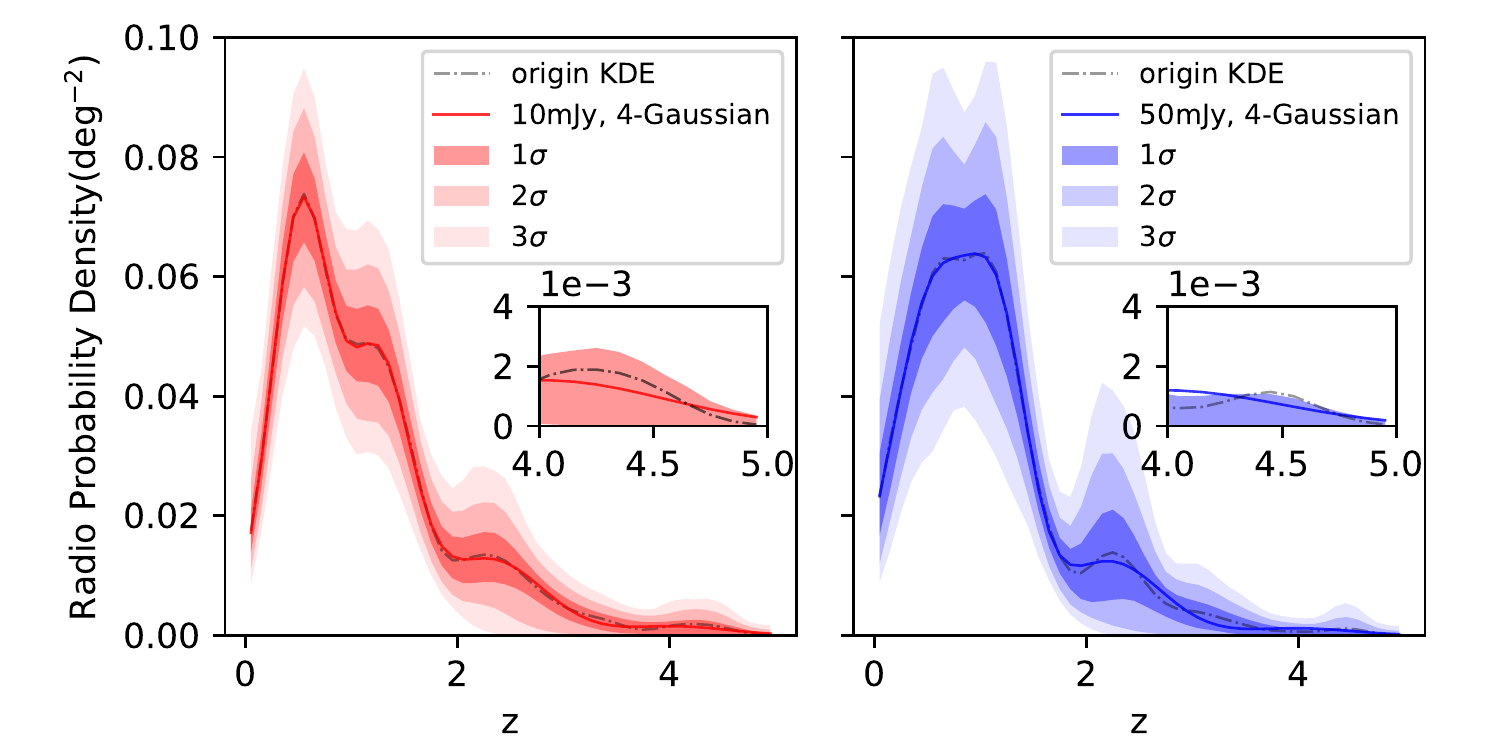}
	\caption{The PDF and 3$\sigma$ errors of 0.2-bandwidth radio KDE curve. The left panel is for $F_\mathrm{1.4GHz}\geq50\mathrm{mJy}$ and the right is for $F_\mathrm{1.4GHz}\geq10\mathrm{mJy}$. The solid and dashed-dotted lines are 4-Gaussian fitted and their original KDE curve. The darkest and lightest areas are the 1$\sigma$ and 3$\sigma$ errors respectively. The inserted small panels zoom in their own tails in the high-$z$ region, showing the imperfection of the fourth component.}
	\label{fig7c}
\end{figure}

We re-sample the three radio datasets 10000 times separately, make number-weighted 0.2-bandwidth KDE, extract its 3$\sigma$ range in every abscissa ($z$) point, and plot them in Figure \ref{fig7c}. The bootstrap 1$\sigma$ errors are more inclusive than the parameter-fluctuation errors in multi-Gaussian (Table \ref{tab3}), and used by us.

The above procedure gives the radio source $z$ probability density with different $F_\mathrm{1.4GHz}$ levels, which needs to multiply with a number density per square degree sky $\sigma$ to become number density. With number-weighted average from the three datasets, we obtain  $\sigma\prime_\mathrm{50}=(105^2/29.2+25^2/6+11^2/2)/(105+25+11)\approx3.714$ and $\sigma\prime_\mathrm{10}$ is 16.966. Similar to \cite{allison22pasa}, we also search the $\sigma$ in the whole NVSS catalogue with a max radio diameter filter ($\theta$<60arcsec) to exclude the extended BRSs, which worsens the covering factor of DLA. Given that the NVSS covers 82\% of the entire sky (33830 $\mathrm{deg^2}$), we collect $N_\mathrm{50mJy}$=125525 and $N_\mathrm{10mJy}$=544130, finally adopting $\sigma_\mathrm{50}$=3.71 and $\sigma_\mathrm{10}$=16.09 instead of $\sigma\prime_\mathrm{50}$ and $\sigma\prime_\mathrm{10}$.

\subsection{Forecasting of observed H21As}\label{sec3.2}
The recent research of DLA distribution at wide $z$ region ($0\lesssim z\lesssim5$) was advanced by \cite{rao17mn}. They provided an optical MgII preselection criterion of DLAs, used it to identify 70 DLAs at low $z$ region ($z\lesssim1.65$) from 369 MgII absorbers, and combined many high-$z$ ($z\gtrsim1.65$) SDSS-detected DLAs and a low-$z$ modified value to give a global redshift number density:
\begin{equation}\label{eq11}
n_\mathrm{Dglo}(z)=(0.027\pm0.007)(1+z)^{(1.682\pm0.200)}.
\end{equation}

\begin{figure}
	\centering
	\includegraphics[scale=0.56]{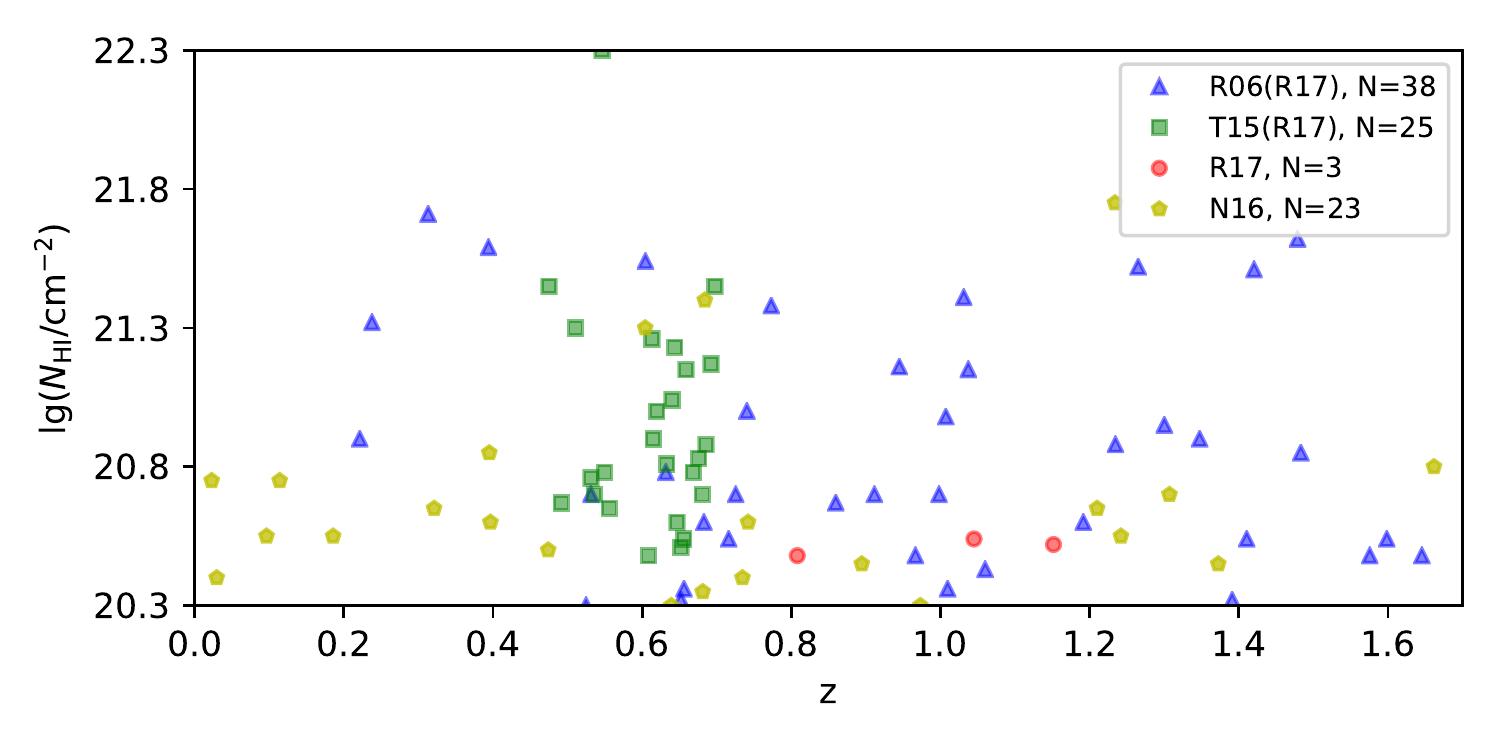}
	\caption{The distribution of 87 DLAs for number density $n_\mathrm{D}$. The blue triangles, green squares and red circles from R06, T15 and R17 separately. The three datasets were all included in R17 article. The yellow pentagons come from N16. Their sample amounts are listed in the legend.}
	\label{fig8}
\end{figure}

Although their result is enough for the $z$ evolution of DLAs, they smoothed out the local DLA distribution features by three wide $z$ bins, which is inadequate for cosmic acceleration observation. So we use their identified 64 DLAs ($N_\mathrm{HI}>2\times10^{20}\mathrm{cm^{-2}}$) from 369 MgII absorbers, and miss 6 DLAs from their full samples. First, \cite{rao06apj} presented 41 DLAs from 197 MgII absorbers, while we only find 38 DLAs with $W_\mathrm{0}^{\lambda2796}>0.6\mathrm{\AA}$. Then, \cite{turns15mn} selected 26 DLAs from 96 MgII absorbers, while we just extract 23 DLAs. \cite{neele16apj} provided extra 23 DLAs with $z$ and $N_\mathrm{HI}$. We plot all 87 DLAs in Figure \ref{fig8}.

\begin{figure}
	\centering
	\includegraphics[scale=0.56]{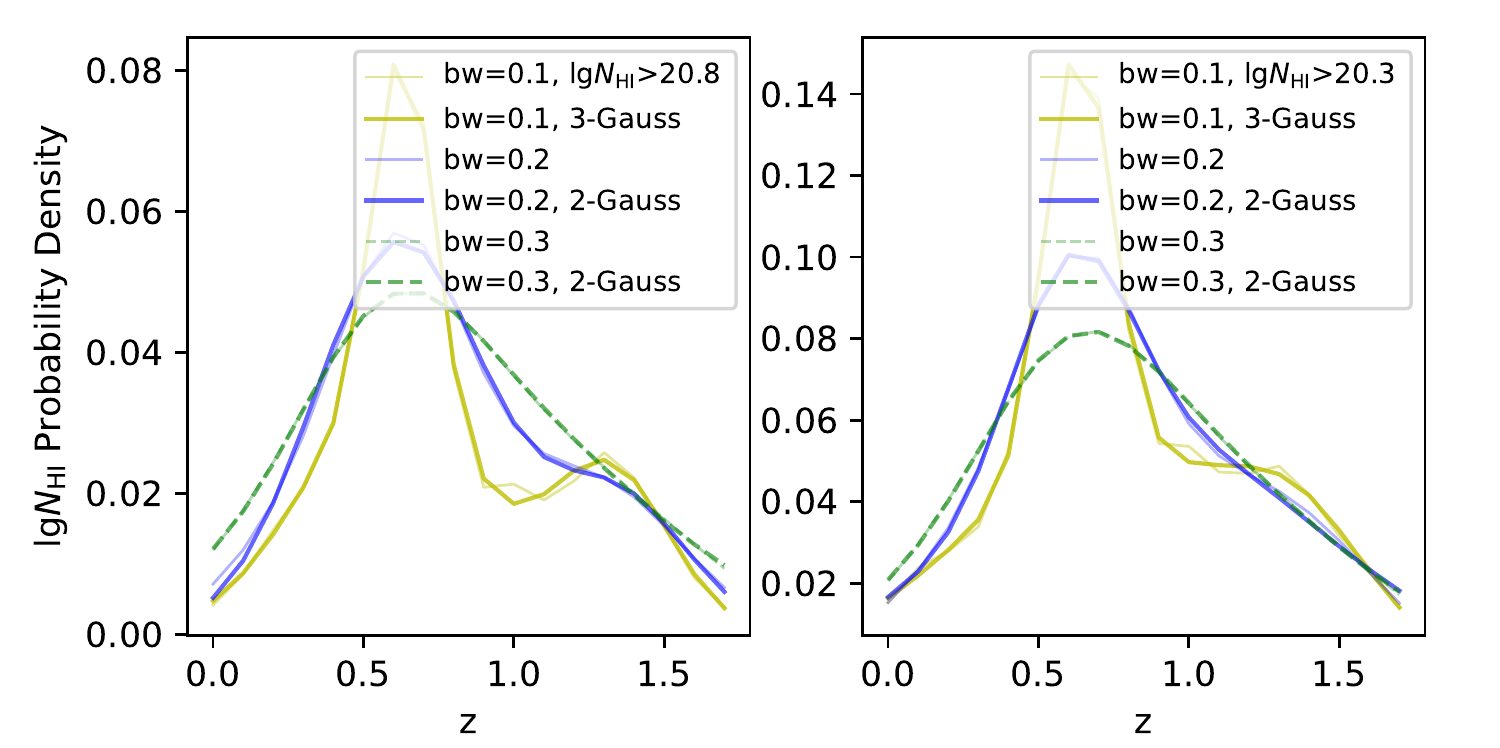}
	\caption{The multi-Gaussian fittings of KDE curves for 87 DLAs with different bandwidths. The left panel is for $\lg N_\mathrm{HI}\geq20.8$ and the right is for $\lg N_\mathrm{HI}\geq20.3$. The thin pale curves are KDE curves with different bandwidths, and the thick dark lines are their fittings respectively.}
	\label{fig9}
\end{figure}

\begin{figure}
	\centering
	\includegraphics[scale=0.56]{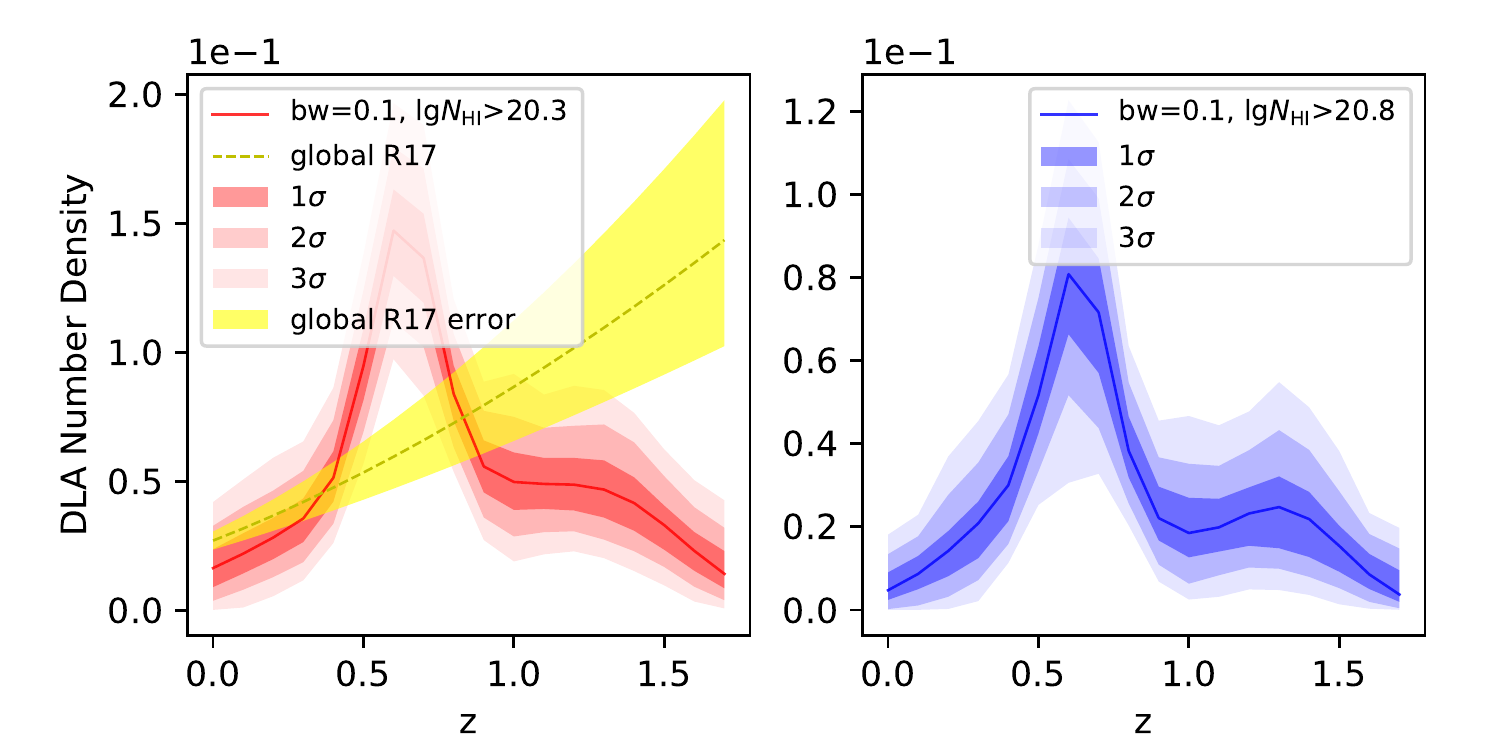}
	\caption{The number distribution and 3$\sigma$ errors of 87 DLAs. The left panel is for $\lg N_\mathrm{HI}\geq20.3$ and the right is for $\lg N_\mathrm{HI}\geq20.8$. The solid lines are their 3-Gauss fitting of KDE curves, and the colored areas present the 3$\sigma$ regions individually. The yellow dashed line and area in the left panel are the DLA number distribution and error region from R17.}
	\label{fig10}
\end{figure}

Similarly, we make 2-dimension ($z$-$N_\mathrm{HI}$) KDE for the 87 DLAs with two lower $N_\mathrm{HI}$ levels, $N_\mathrm{HI}>2\times10^{20}\mathrm{cm^{-2}}$ ($\lg N_\mathrm{HI}>20.3$ for visual simplicity, although it should be $\lg (N_\mathrm{HI}/\mathrm{cm^{-2}})>20.3$) and $N_\mathrm{HI}>6.31\times10^{20}\mathrm{cm^{-2}}$ ($\lg N_\mathrm{HI}>20.8$), and use manual K-S test to compare the constant C and their maxVDs. Provided the simple profiles of KDE in Figure \ref{fig9}, we choose a 3-Gaussian fitting with 0.1 bandwidth for accuracy, and list the fitted values and errors in Table \ref{tab4}. A 10000-bootstrap gives the 1$\sigma$ errors in Figure \ref{fig10}.
\begin{table}
	\centering
	\caption{Multi-Gaussian fitting of 0.1-bandwidth DLA KDE curves}\label{tab4}
		\begin{tabular}{cccc}
			\toprule
			component & a & $\mu$ & $\sigma$\\ 
			\midrule
			20.8-1 & 0.0323$\pm$0.0016 & 0.5789$\pm$0.0097 & 0.2953$\pm$0.0098\\
			20.8-2 & 0.0517$\pm$0.0015 & 0.6350$\pm$0.0020 & 0.0993$\pm$0.0028\\
			20.8-3 & 0.0231$\pm$0.0006 & 1.3149$\pm$0.0084 & 0.2007$\pm$0.0077\\
			
			20.3-1 & 0.1025$\pm$0.0021 & 0.6336$\pm$0.0018 & 0.1099$\pm$0.0024\\
			20.3-2 & 0.0509$\pm$0.0024 & 0.7506$\pm$0.0495 & 0.4782$\pm$0.0388\\
			20.3-3 & 0.0233$\pm$0.0059 & 1.3649$\pm$0.0197 & 0.2273$\pm$0.0315\\
			\bottomrule
		\end{tabular}
\end{table}

With the 21cm absorption optical depth $\tau_\mathrm{21}(\nu)=-\log\{1-\Delta F(\nu)/[C_\mathrm{f}F(\nu)]\}$, $N_\mathrm{HI}$ can be expressed as \citep{kanek14mn}:
\begin{equation}\label{eq12}
N_\mathrm{HI}=1.823\times10^{18}T_\mathrm{S}\int-\log[1-\frac{\Delta F(\nu)}{C_\mathrm{f}F(\nu)}]d\nu,
\end{equation}
where $T_\mathrm{S}$ and $C_\mathrm{f}$ is spin temperature and covering factor of DLA, $\Delta F$ is the absorbed flux density, and $F$ is the continuum flux density. Because of the low 21cm optical depth ($\tau_\mathrm{21}\ll1$) for most DLAs, eq. \ref{eq12} can reduce to:
\begin{equation}\label{eq13}
	N_\mathrm{HI}=1.823\times10^{18}\frac{T_\mathrm{S}}{C_\mathrm{f}}\int\frac{\Delta F(\nu)}{F(\nu)}d\nu.
\end{equation}
$N_\mathrm{HI}$ is determined by $T_\mathrm{S}$, $C_\mathrm{f}$ and integrated percent absorption. Constraints on $N_\mathrm{HI}$ can be decomposed into three parts.

However, according to the same article \citep{kanek14mn}, no statistical relations are found in $T_\mathrm{S}$-$N_\mathrm{HI}$ and $T_\mathrm{S}$-$C_\mathrm{f}$, while $N_\mathrm{HI}$-$\int\tau_\mathrm{21}(\nu)d\nu$ has a $3.8\sigma$-significance correlation. They find a 37-DLA dataset with their $z$, $N_\mathrm{HI}$, estimated $T_\mathrm{S}$ and $C_\mathrm{f}$ in the foreground of compact radio quasars. Depending on this dataset, we will make a 3-dimension KDE in the parameter space of $z$-$N_\mathrm{HI}$-$T_\mathrm{S}$, and limit $N_\mathrm{HI}$ and $T_\mathrm{S}$ as before. Because the $N_\mathrm{HI}$-$C_\mathrm{f}$ relation was not explored in their paper, we do not limit $C_\mathrm{f}$ together.

\cite{grasha20mn} found the best $T_\mathrm{S}$ is 175/$C_\mathrm{f}$K for low-$z$ (z<1) DLAs, and 100/$C_\mathrm{f}$-500/$C_\mathrm{f}$K is acceptable. \cite{allison21mn} obtained 274K and 576K in the low-$z$ region (z<1)  in the detection and non-detection situations (2$\sigma$). Overall, we take $T_\mathrm{S}$ of 500K and 1000K as two limits, with the former two $\lg N_\mathrm{HI}$ limits (20.3, 20.8) in our estimation of $\kappa(z,\lg N_\mathrm{HI},T_\mathrm{S})$.

\begin{figure}
	\centering
	\includegraphics[scale=0.56]{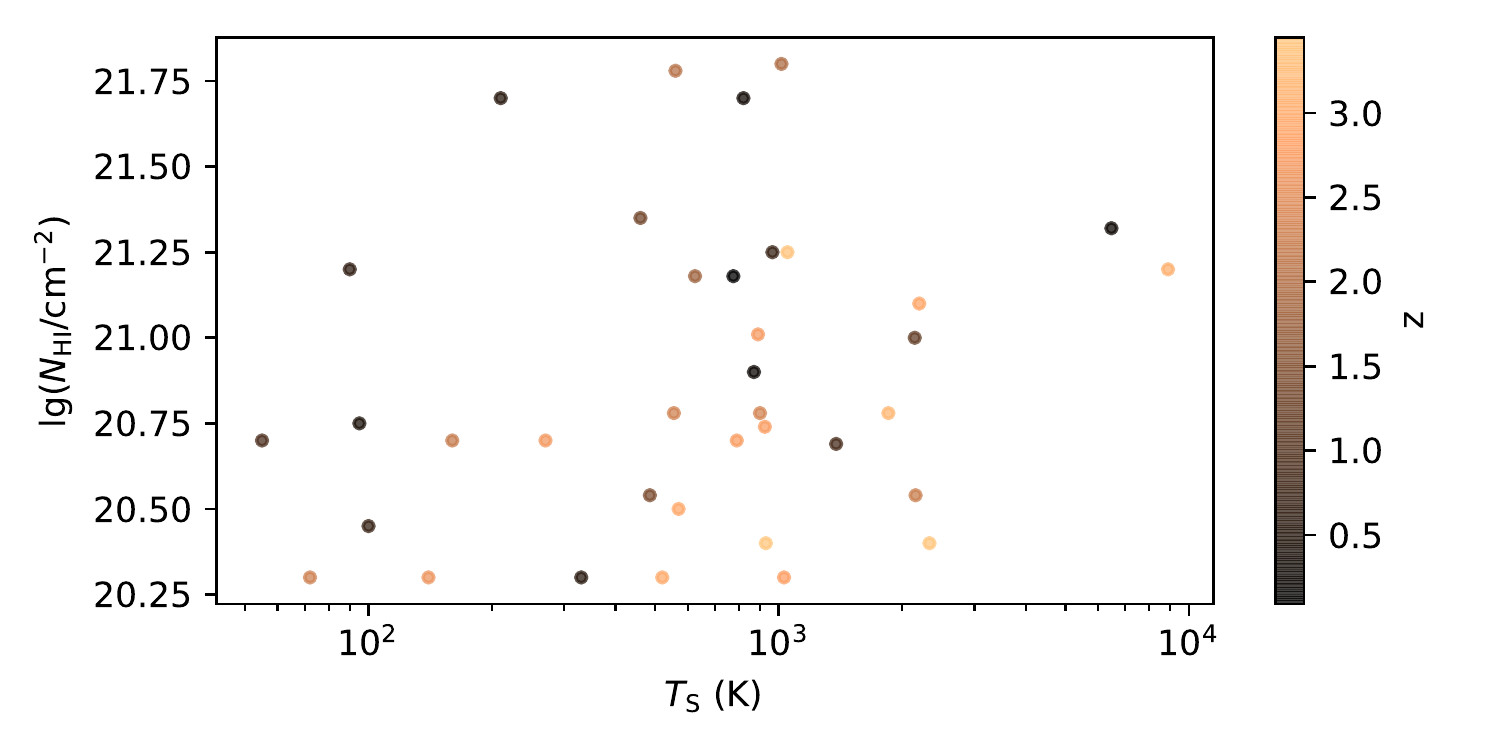}
	\caption{The distribution of 37 DLAs for detection rate $\kappa$. The darker sample has a larger redshift.}
	\label{fig10a}
\end{figure}

\begin{figure}
	\centering
	\includegraphics[scale=0.56]{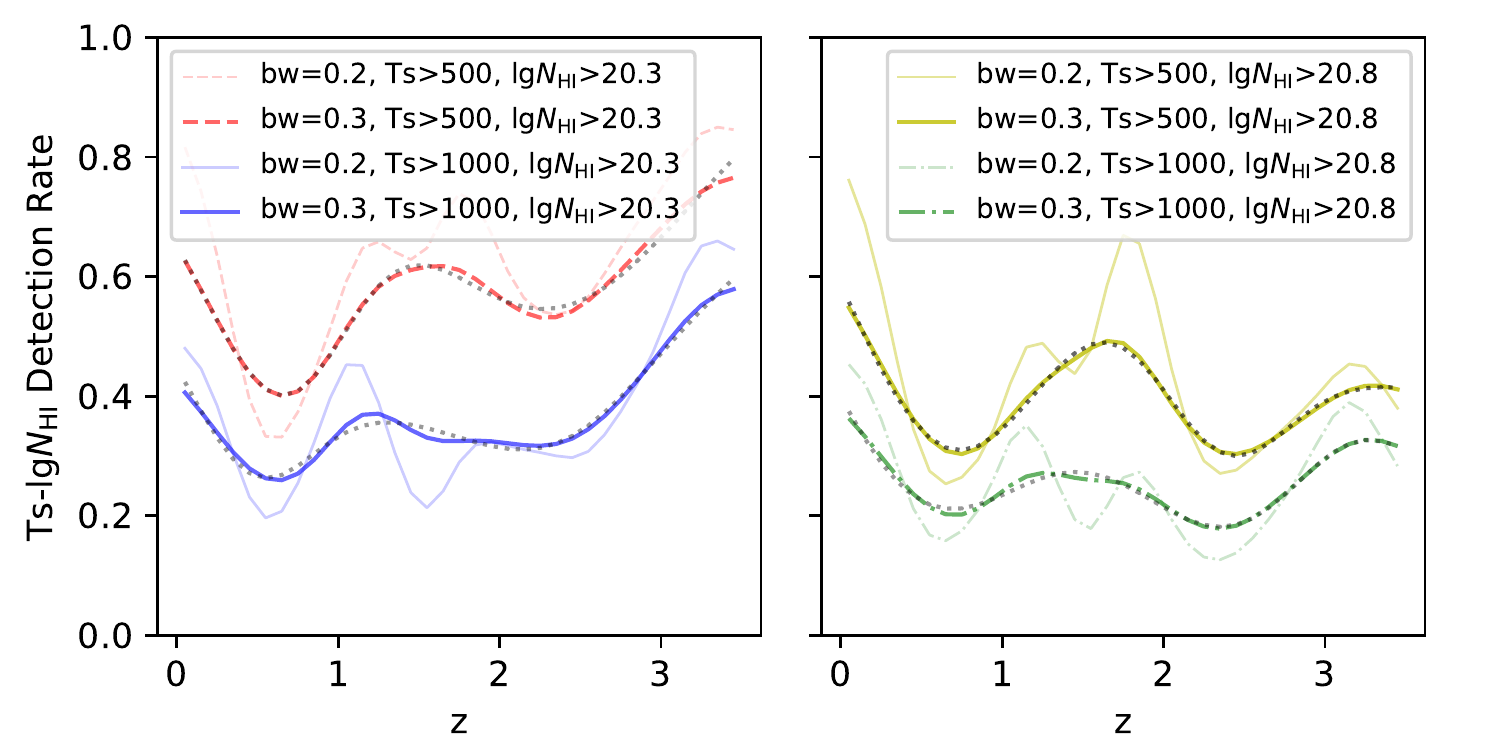}
	\caption{The multi-Gaussian fittings of KDE curves for 37 DLAs with different bandwidths. The left panel is for $\lg N_\mathrm{HI}\geq20.3$ and the right is for $\lg N_\mathrm{HI}\geq20.8$. The thin pale curves are 0.2-bandwidth KDE curves and the thick dark lines are 0.3-bandwidth ones. The unlabeled gray dotted lines are 3-Gaussian fitting for 0.3-bandwidth KDEs.}
	\label{fig10b}
\end{figure}

\begin{figure}
	\centering
	\includegraphics[scale=0.58]{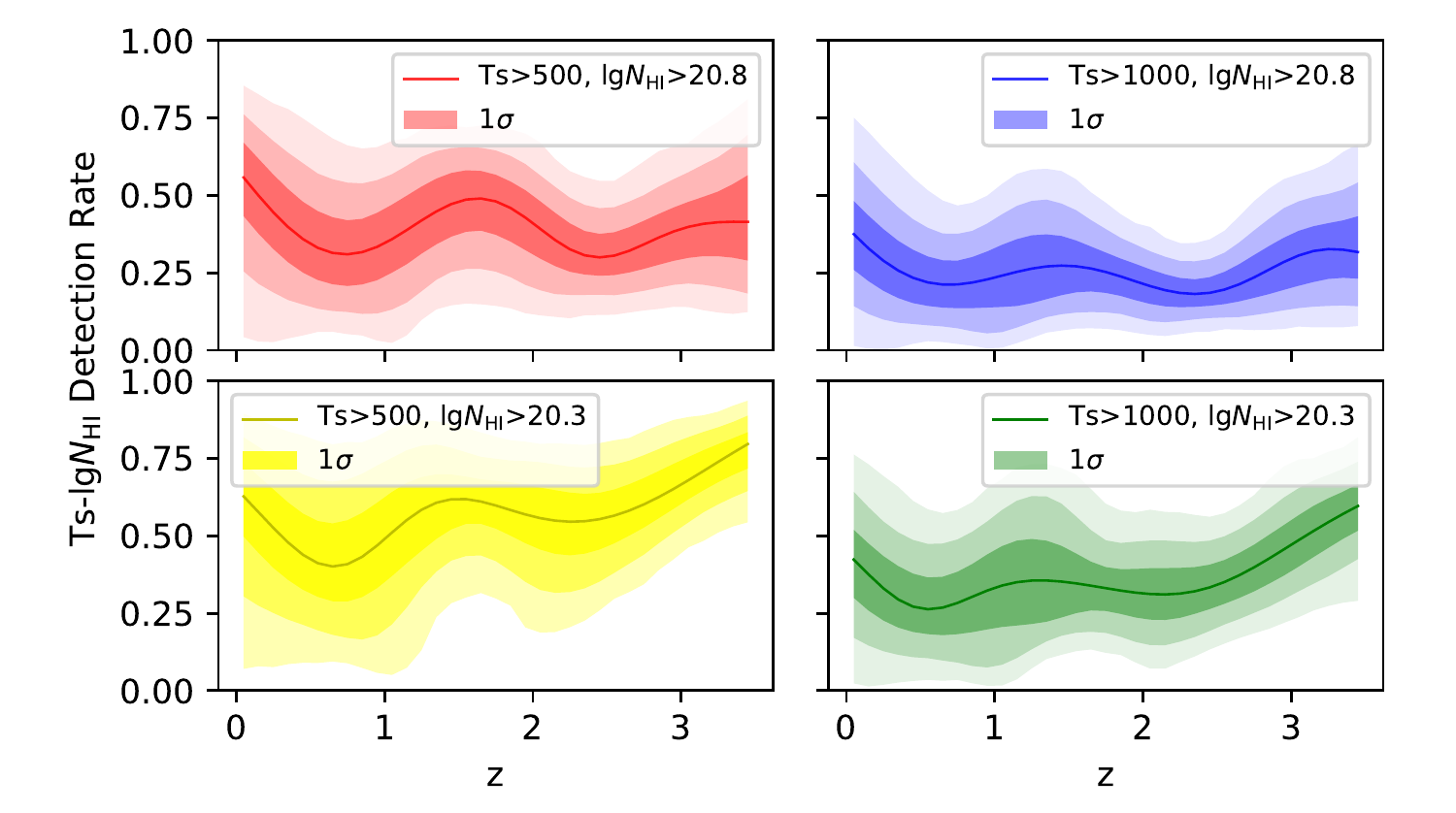}
	\caption{The detection rates and 3$\sigma$ errors of 37 DLAs. The left panels are for $T_\mathrm{S}$>500K and the right are for $T_\mathrm{S}$>1000K. The upper panels are for $\lg N_\mathrm{HI}\geq20.8$ and the lower are for $\lg N_\mathrm{HI}\geq20.3$. The solid lines are the 3-Gauss fitting of KDEs, and the colored areas present their 3$\sigma$ errors individually. Only the darkest 1$\sigma$ regions are labeled in the legends.}
	\label{fig10c}
\end{figure}

Again, we plot the extra 37 DLAs in Figure \ref{fig10a}. Making 3-dimension KDE with $T_\mathrm{S}$ and $\lg N_\mathrm{HI}$ limits and conducting a manual K-S test, we perform a 3-Gaussian fitting (considering the actual shape of detection rate curves, we use one subtract the 3-Gaussian as the results) with 0.3 bandwidth in Figure \ref{fig10b} with their parameters listed in Table \ref{tab5}. The less bandwidth and more Gaussians are discarded to alleviate overfitting. The 3$\sigma$ errors from 10000-bootstrap are offered in Figure \ref{fig10c}. The 20.3-500 situation in Table \ref{tab5} has the most unsuccessful constraint, reflected by the widest 3$\sigma$ region in Figure \ref{fig10c}. But considering the good performance between the KDE curve and its multi-Gaussian fitting, we still accept the results.
\begin{table}
	\centering
	\caption{Multi-Gaussian fitting$^1$ of 0.2-bandwidth detection KDE curves}\label{tab5}
	\begin{tabular}{cccc}
		\toprule
		component & a & $\mu$ & $\sigma$\\ 
		\midrule
		20.8-1000-1 & 0.4686$\pm$0.1408 & 3.8692$\pm$0.1728 & 0.4394$\pm$0.0960\\
		20.8-1000-2 & 0.7740$\pm$0.0154 & 2.4748$\pm$0.0154 & 0.8251$\pm$0.0534\\
		20.8-1000-3 & 0.7309$\pm$0.0174 & 0.5183$\pm$0.0324 & 0.7968$\pm$0.0282\\
		
		20.3-1000-1 & 0.5850$\pm$0.0084 & 3.6992$\pm$0.1645 & 1.6574$\pm$0.2301\\
		20.3-1000-2 & 0.2529$\pm$0.0285 & 2.3546$\pm$0.0097 & 0.4074$\pm$0.0180\\
		20.3-1000-3 & 0.5783$\pm$0.0499 & 0.6383$\pm$0.0187 & 0.6637$\pm$0.0254\\
		
		20.8-500-1 & 0.1109$\pm$0.0156 & 2.4000$\pm$0.0280 & 0.5282$\pm$0.0506\\
		20.8-500-2 & 0.1890$\pm$0.0304 & 0.4603$\pm$0.0331 & 0.3654$\pm$0.0380\\
		20.8-500-3 & 0.6210$\pm$0.0126 & 1.5104$\pm$0.1410 & 2.0000$\pm$0.1354\\
		
		20.3-500-1 & 0.2378$\pm$1.1207 & 0.0602$\pm$3.4411 & 0.5034$\pm$2.3483\\
		20.3-500-2 & 0.4541$\pm$0.0020 & 2.2745$\pm$0.0136 & 0.9268$\pm$0.0163\\
		20.3-500-3 & 0.3842$\pm$2.2242 & 0.6983$\pm$0.7163 & 0.4088$\pm$0.2259\\
		\bottomrule
	\end{tabular}
	\begin{tablenotes}
		\footnotesize
		\item{$^1$} The actual curve of detection rate is the result that one subtracts the 3-Gaussian fitting.
	\end{tablenotes}
\end{table}

\subsection{Anticipation of potential H21As}\label{sec3.3}
\begin{figure}
	\centering
	\includegraphics[scale=0.58]{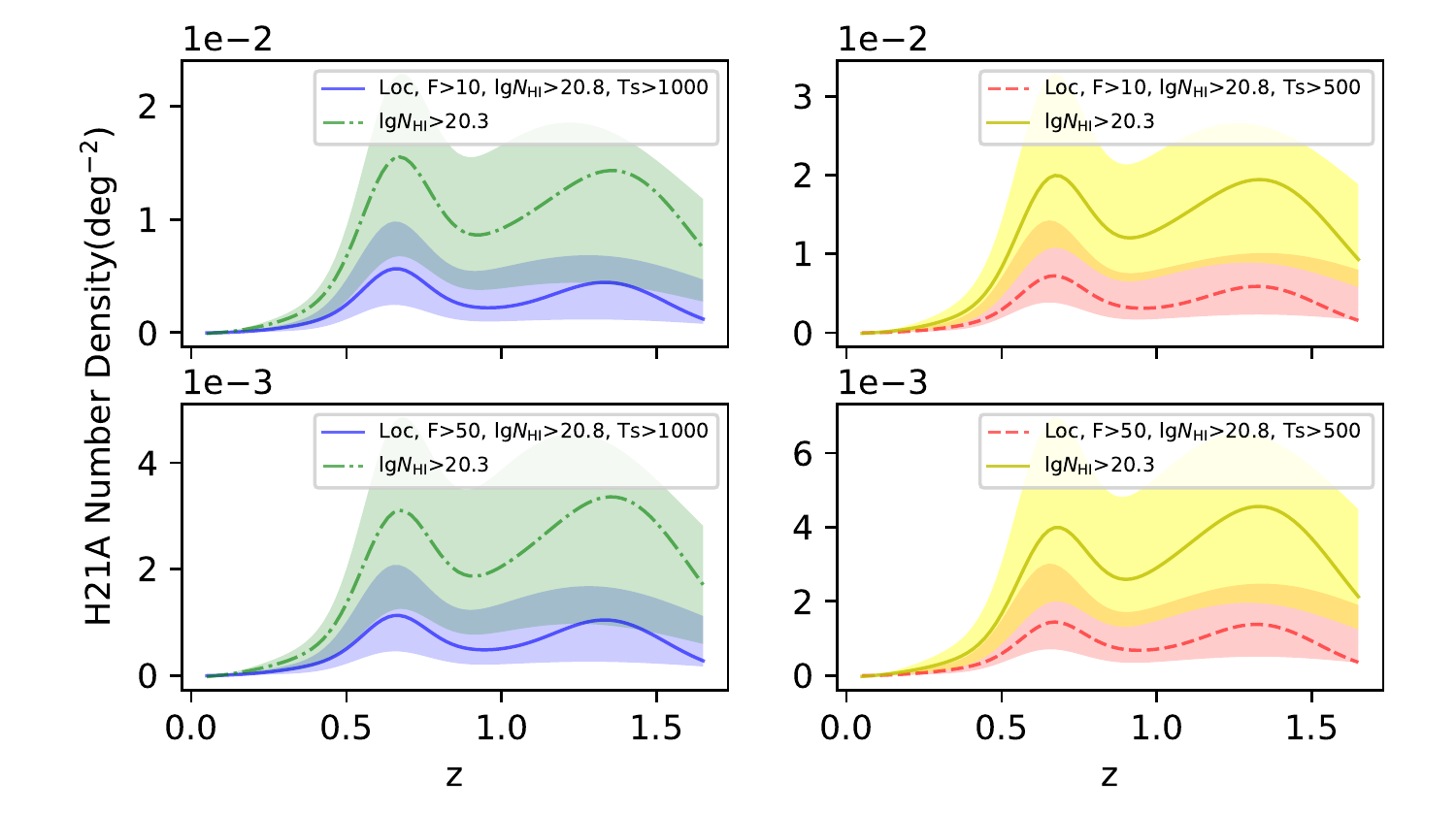}
	\caption{The potential H21A number density. The left panels are for $T_\mathrm{S}$>1000K and the right are for $T_\mathrm{S}$>500K. The upper panels are for $F_\mathrm{1.4GHz}$>10mJy and the lower are for $F_\mathrm{1.4GHz}$>50mJy. The lines and corresponding color areas are their density curves and 1$\sigma$ errors.}
	\label{fig11}
\end{figure}

\begin{figure}
	\centering
	\includegraphics[scale=0.58]{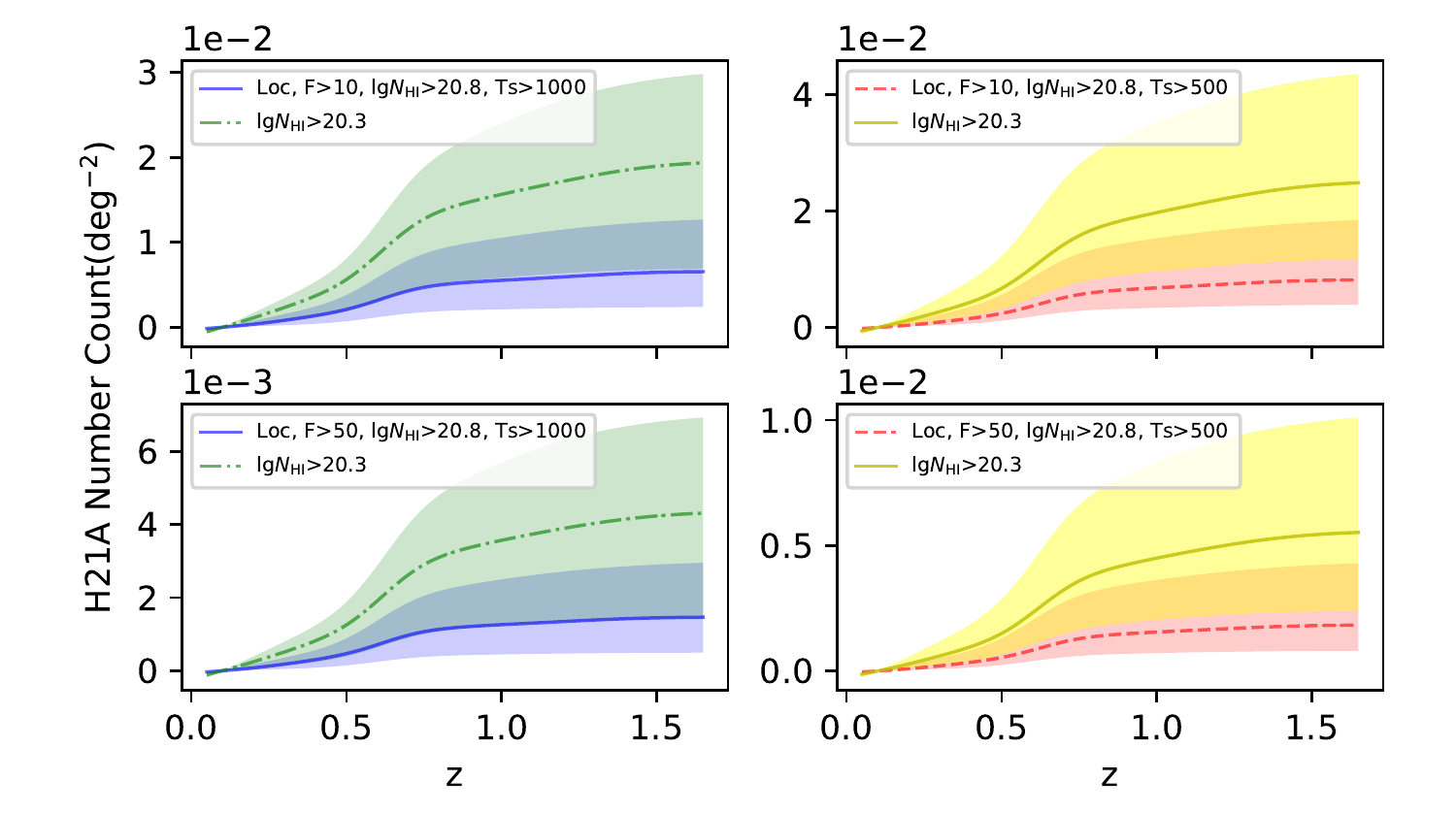}
	\caption{The potential H21A number count. The left panels are for $T_\mathrm{S}$>1000K and the right are for $T_\mathrm{S}$>500K. The upper panels are for $F_\mathrm{1.4GHz}$>10mJy and the lower are for $F_\mathrm{1.4GHz}$>50mJy. The lines and corresponding color areas are their count curves and 1$\sigma$ errors.}
	\label{fig12}
\end{figure}

Given all the components, we can calculate the potential H21As number density (eq. \ref{eq8}) in Figure \ref{fig11} and its integrated number count in Figure \ref{fig12} from z=0.1 to eliminate the objects in the local Universe.

FAST covers about 24000$\mathrm{deg^2}$ \citep{li18imm} and the redshift of 0 to 0.352 (1050-1450MHz). ASKAP covers 33000$\mathrm{deg^2}$ \citep{allison22pasa} and the redshift of 0.4 to 1.0 (711.5-999.5MHz). As for the most remarkable instrument, the SKA1-Mid (SKA1M) covers about 30000 $\mathrm{deg^2}$ \citep{klo15aaska,welt20pasa} and the redshift of 0 to 1. We list the predicted detection yield in Table \ref{tab6} for each instrument and lower limit condition, and the 1$\sigma$ errors are multiplicative result from the three-component bootstraps.
\begin{table}
	\centering
	\caption{The detection yield of potential H21As in each condition}\label{tab6}
	\begin{tabular}{ccccc}
		\toprule
		Telescope & $F_\mathrm{min}$/mJy & $\lg(N_\mathrm{HI,min}$/$\mathrm{cm^{-2}})$ & $T_\mathrm{S,min}$/K & Prediction \\ 
		\midrule
		FAST & >50 & >20.8 & >1000 & $5.8^{+5.2}_{-4.2}$\\
		 &  &  & >500 & $6.6^{+9.4}_{-4.0}$\\
		 &  & >20.3 & >1000 & $16.0^{+8.3}_{-11.4}$\\
		 &  &  & >500 & $18.7^{+17.8}_{-9.8}$\\
		FAST & >10 & >20.8 & >1000 & $26.2^{+21.4}_{-18.6}$\\
		 &  &  & >500 & $30.0^{+39.4}_{-17.2}$\\
		 &  & >20.3 & >1000 & $72.1^{+33.3}_{-49.7}$\\
		 &  &  & >500 & $84.7^{+73.2}_{-46.0}$\\
		ASKAP & >50 & >20.8 & >1000 & $31.5^{+32.2}_{-19.6}$\\
		 &  &  & >500 & $39.8^{+52.7}_{-20.8}$\\
		 &  & >20.3 & >1000 & $90.7^{+56.1}_{-57.8}$\\
		 &  &  & >500 & $116.5^{+96.6}_{-63.0}$\\
		ASKAP & >10 & >20.8 & >1000 & $137.2^{+129.8}_{-81.1}$\\
		 &  &  & >500 & $172.8^{+214.7}_{-83.6}$\\
		 &  & >20.3 & >1000 & $394.2^{+220.6}_{-250.1}$\\
		 &  &  & >500 & $506.7^{+385.6}_{-255.3}$\\
		SKA1M & >50 & >20.8 & >1000 & $37.8^{+37.1}_{-24.8}$\\
		 &  &  & >500 & $46.6^{+62.3}_{-25.1}$\\
		 &  & >20.3 & >1000 & $107.0^{+63.7}_{-69.9}$\\
		 &  &  & >500 & $135.0^{+114.7}_{-73.9}$\\
		SKA1M & >10 & >20.8 & >1000 & $165.8^{+150.2}_{-102.5}$\\
		 &  &  & >500 & $204.2^{+255.3}_{-102.7}$\\
		 &  & >20.3 & >1000 & $469.2^{+250.4}_{-293.8}$\\
		 &  &  & >500 & $591.7^{+461.7}_{-302.9}$\\
		\bottomrule
	\end{tabular}
\end{table}

\section{discussion}\label{sec4}
We emphasise that the S-L signal is a direct measurement, and its theoretical derivation is independent of Einstein's equation and Copernican principle \citep{yu14prl}, compared with other common cosmological probes like SNe Ia. But the different matter structures of every sightline in the signal propagation necessitates further studies of the inhomogeneity influence on the S-L signal.

The systematics of the S-L effect mainly contains two parts. First, the proper acceleration of the observer involves the earthly, solar, galactic and cosmological frames. The former two terms can be calculated in the highest level of cm/s (one-decade accumulation of S-L signal) \citep{wirght14pasp}. The solar revolution around the galactic center can be accurately measured by pulsar timing \citep{zakam05aj}, Gaia Data Release 3 \citep{gaia21aap} and VLBI \citep{xu12aa,titov18aap} in mm/s/yr level which is the same as S-L signal. And the proper motion of extragalaxies is accessible by VLBI \citep{titov11aa,titov22mn}. The possible redshift uncertainties were given in \cite{bolej19arx} and reviewed in \cite{lu21arx}. Second, the peculiar motion of absorption line systems would impact the S-L signal. Combining the various physical conditions of DLAs \citep{dutta17mn}, the Neutral Mass (NM) ratio (Cold NM:Unstable NM:Warm NM = 28:20:52) in our galaxy \citep{murray18apjs}, and Figure \ref{fig10a} from \cite{kanek14mn}, there do exist some DLAs with sharp absorptions as ideal targets for probing S-L signal. Additionally, the data process introduces new uncertainty when signal extraction and flux calibration. It is not the primary challenge but needs to maintain the same procedure and reform with new techniques.

Though simple statistical models like exponential can determine the number densities, and KDE would decrease generalization ability. We still use KDE for some reasons, most importantly to keep the method consistent in the data process. Second, as a non-parametric estimation, KDE needs less prior knowledge about the realistic distribution. Third, these simple-peak models are inadequate for multi-peak distribution. Fourth, when filtering data with different lower limits, some samples are discarded but close to the limits. However, KDE will record the contributions of these border-near samples. Although KDE is a sensitive method suffering more fluctuations from many factors, it provides similar features (a big bulge containing two peaks and a small bulge/peak) in the two panels of Figure \ref{fig7c}, proving its robustness sufficiently.

The multi-Gaussian fittings for 2 components of KDEs in Table \ref{tab3} and \ref{tab5} offer poor constraints, but the fittings quite approach the KDEs and such a defect does not hinder our estimation.

Most used radio samples are radio-loud AGNs, but other information such as morphological and spectral types was not recorded. One can better grip their redshift distribution and radio luminosity function by dividing samples into more sub-types if acquiring more data, but now they are beyond our present reach.

\begin{figure}
	\centering
	\includegraphics[scale=0.58]{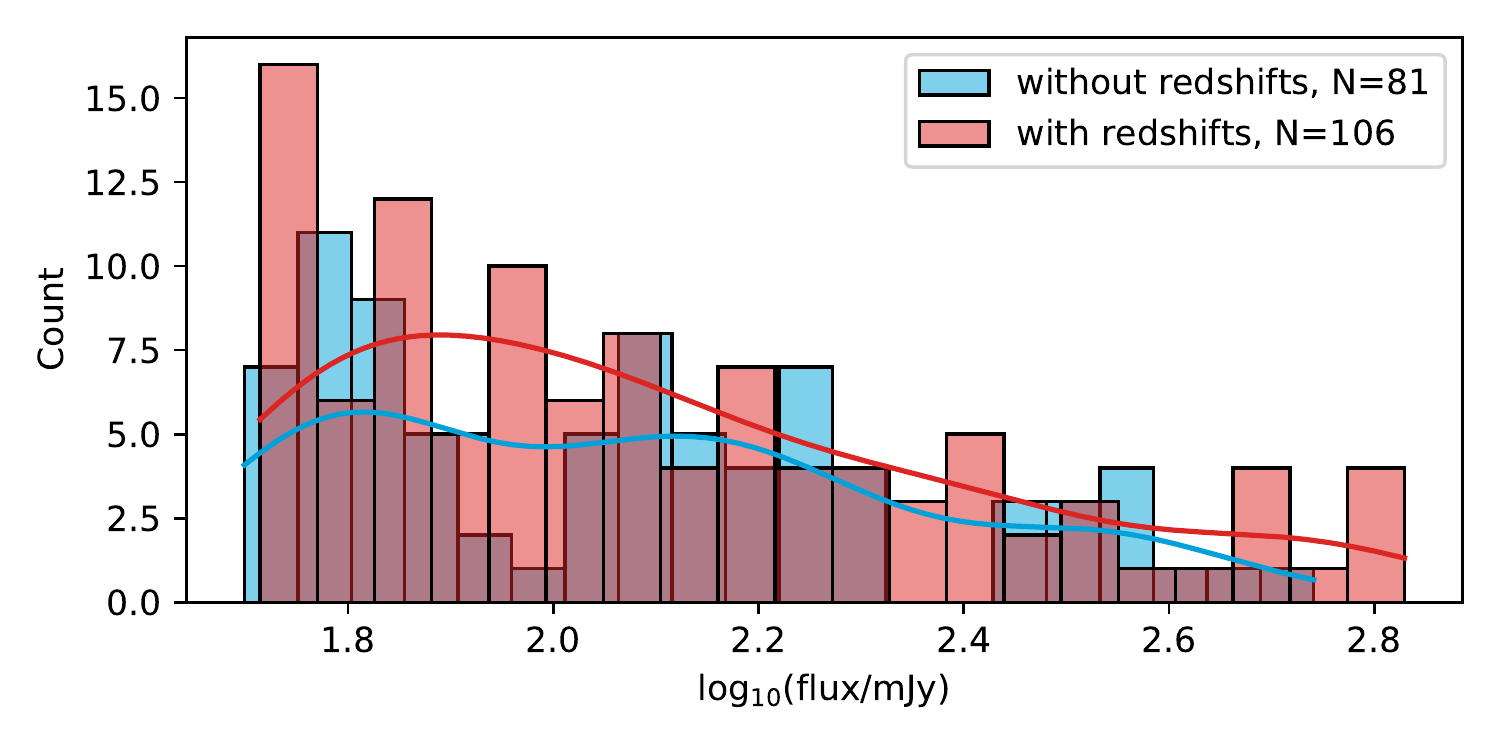}
	\caption{The histogram of CoNFIG-4 data for $F_\mathrm{1.4GHz}$. The blue bars are for the samples without redshifts, and the blue line is their KDE curve, while the red ones are for those redshift "owners".}
	\label{fig12a}
\end{figure}

With different radio datasets, it is necessary to compare their observations. The radio sources in CENSORS and Hercules were preliminarily selected by pioneering radio mapping and then probed by subsequent optical spectra. Original CENSORS \citep{brook08mn} optically observed 143 in all 150 sources, providing a 71\% spectroscopic completeness. As for the rest, the $K$-$z$ (and $I$-$z$ for one source) relation was used to estimate their redshifts. Original Hercules \citep{waddi01mn} with 72 sources, offered 47 spectroscopic redshifts, found 10 upper limits of redshift from literature, and estimated the broad-band photometric redshifts for the rest. \cite{rigby11mn} rearranged the two sets with some published reassessments. They selected a 135-sample subset for CENSORS with 73\% spectroscopic completeness, and a 64-sample subset for Hercules with 40 spectroscopic and 20 photometric redshifts. As for CoNFIG-4 samples \citep{gendre09mn,gendre10mn}, they were selected from NVSS and FIRST \citep{white97apj} counterplots with complete Fanaroff–Riley morphology, and reviewed for their optical information. No observation was performed in CoNFIG papers, hence their redshift information was all from literature. For the samples without redshifts in CoNFIG-4, they have roughly similar but lower flux distribution compared to redshift "owners" in Figure \ref{fig12a}, so it is hard to regard them as statistically less luminous in $F_\mathrm{1.4GHz}$. But we find that most samples without redshifts do not have SDSS magnitudes recorded in the original data.

Optical selections inserted in observation unavoidably bias radio number density, but more datasets can provide a more robust prediction than a single one. Like the radio source number density per square degree sky ($\sigma$) at the end of section \ref{sec3.1.3}, only with three datasets, our estimated $\sigma_{50}\prime$=3.714 can approach the NVSS value $\sigma_{50}$=3.71, while any single set corresponds a greater deviation.

\cite{rao17mn} provided abundant DLAs at low-redshift ($z\lesssim1.65$) where the ground-based optical telescope can not directly observe. A reasonable speculation is that their preselection is biased by luminosity and dust extinction, having the risk of omitting some MgII absorbers, DLAs without MgII absorption, and deriving biasedly higher $\Omega_\mathrm{DLA}$. However, they found (in sec 4 of that article) no evidence for these effects at $z$<1.65. And Kanekar's conclusion\citep{kanek14mn} also disapproves of this selection effect in the aspect of $T_\mathrm{S}$ distributions.

We notice that \cite{dutta17mn,dutta19jaa} advanced an absorption-blind or galaxy-selected approach, selecting a galaxy visually close to a background quasar (Quasar-Galaxy Pair, QGP) without the prior of any absorption toward the quasar. Their method is not intuitively biased as MgII preselection. We hope their project finds more QGPs for the distribution of radio DLASs.

\begin{figure}
	\centering
	\includegraphics[scale=0.58]{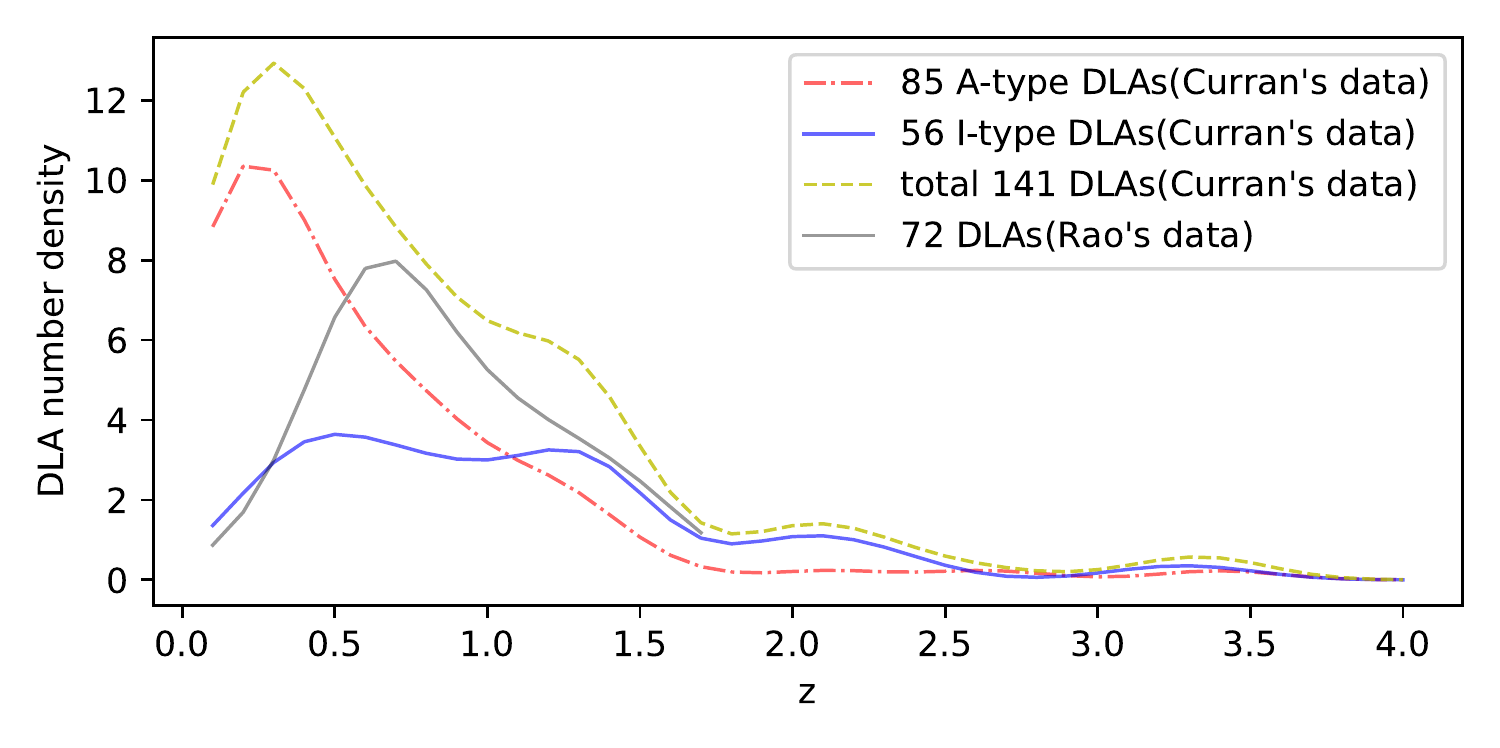}
	\caption{The KDE fittings (bandwidth=0.2) of two DLA datasets. The gray and yellow dashed curves are for the total two sets. The red dashed-dotted and blue curves are the A- and I-type DLAs in Curran's data separately.}
	\label{fig13}
\end{figure}

In addition to Rao's and Neeleman's I-type DLAs, there exist other DLA datasets, e.g.  \cite{curran16mn,curran21mn}. Curran's data provide 85 A-type and 56 I-type DLAs with $z$>0.1. Although Curran's data cover wider redshift and more samples, they are collected from different observations. The decisive reason for discarding Curran's data is they do not contain $N_\mathrm{HI}$ which is included in Rao's data, and we want to use 2-dimension ($z$ and $N_\mathrm{HI}$) KDE for more features. Moreover, we plot Curran's and Rao's data in Figure \ref{fig13}, and Rao's KDE profile of I-type DLAs almost envelopes Curran's, exceptionally raising a bulge mainly because of the subset T15 \citep{turns15mn} with $z$ from 0.4 to 1.0. Besides, it is normal that total Curran's KDE profile covers Rao's, because the former contains extra A-type DLAs which are not suitable for S-L signal. Because over half DLAs (28 in 41) overlap Rao's samples, we do not plot Neeleman's KDE.

The different tendencies in the low-$z$ region between our KDE-derived and Rao's DLA number density in the left panel of Figure \ref{fig10}, apart from their wide $z$-bin, are mainly caused by Rao's modified DLA number density values at z=0 \citep{zwaan05mn,braun12apj}. The modification is reasonable if we consider the left end of Curran's I-type KDE, which is slightly higher than Rao's in Figure \ref{fig13}, indicating the low-$z$ shortage of Rao's samples. But the single value point $n_\mathrm{DLA}(z=0)$ can not change the KDE result unless we have real low redshift DLAs to feed the algorithm.

Our $N_\mathrm{HI}$-$T_\mathrm{S}$ constraints for DLAs and the H21A-to-DLA proportional constant $\eta$, mainly rely on the 37-DLA data from \cite{kanek14mn}. Hence our results are inevitably affected by small samples and Poisson errors, and need more "full" information ($z$, $N_\mathrm{HI}$, $T_\mathrm{S}$ and $C_\mathrm{f}$) DLAs to overcome.

The recent study of DLAs detection yield includes the remarkable work from \cite{allison22pasa}, where they defined a completeness function $\mathscr{C}$ with the 1000 mock spectra recovered from FLASH early survey of the GAMA 23 field \citep{allison20mn}, and used a $N_\mathrm{HI}$ frequency distribution function $\mathscr{F}$ linearly interpolated from z=0 \citep{zwaan05mn} and z>2 \citep{bird17mn}, to predict the number of I-type and A-type DLAs in the ASKAP range. Future researchers will be enlightened to improve their $N_\mathrm{HI}$ frequency distribution. And their completeness function reminds us that more factors, such as signal-to-noise rate and blind-survey resolution (related to H21A's line width distribution), can be considered in prediction. These observational limitations from specific facilities are more concrete, and entail our further efforts.

An interesting difference between our prediction and the Allison's of ASKAP is, in our 37 DLAs, the low-$N_\mathrm{HI}$ I-type ones are more than the high-$N_\mathrm{HI}$. Namely, our higher sensitivity condition in $N_\mathrm{HI}$ (to discover more DLAs) is to lower $N_\mathrm{HI}$, while they need to increase it. Their complicated functions $\mathscr{C}$ and $\mathscr{F}$, both relate to $N_\mathrm{HI}$, and both are not plotted with respect to $N_\mathrm{HI}$. Thus it is hard to say which one ($\mathscr{C}$ or $\mathscr{F}$) dominates this difference.

Given more physical constraints to potential intervening H21As for FAST, our most optimistic expectation is quite small (80), around 1 or 2 order magnitude smaller than some previous work, such as 1500 \citep{zhang21mn} from a local luminosity function estimation and 2600 \citep{jiao20jcap} for a decade CRAFTS observation \citep{zhang19scpma} with a simpler integration of eq. \ref{eq8}. However, our optimistic result approaches the magnitude of 100 from ALFALFA-survey estimation \citep{wu15aas}. Additionally, Zhang's 1500 result may omit that only about 10\% AGNs are radio-loud and ignored AGN's $z$ distribution.

\section{conclusion}\label{sec5}
In this paper, we make a small but important distinction between the global and dynamical $\ddot{a}$ and the local and observed $\dot{v}_\mathrm{S}$ at first, and emphasise that only the $\ddot{a}$ can express an actual expansion state for the whole Universe at one certain time, but it has not been measured model-independently.

Subsequently, in the bulk of the paper, we separately explore (i) $n_\mathrm{R}$, the radio-source redshift number density per square degree via three datasets (with $z$ and $F_\mathrm{1.4GHz}$), (ii) $n_\mathrm{D}$, the DLA redshift number density in the sightline through 87 DLAs from two low-redshift ($z\lesssim1.65$) datasets (with $z$, $N_\mathrm{HI}$), and (iii) $\kappa$, the DLA detection rate at given $N_\mathrm{HI}$-$T_\mathrm{S}$ limitations from an extra 37-DLA dataset and $\eta$ (the fraction of H21As in DLAs). (1) Introducing the KDE method, we replace the traditional description of the $n_\mathrm{R}$ and $n_\mathrm{D}$, and propose a data-based $\kappa$ function, with multi-Gaussian profiles (Table \ref{tab3}, \ref{tab4} and \ref{tab5}) of KDE curves. (2) Adopting the bootstrap method, we give the former three functions 1$\sigma$ errors (Figure \ref{fig7c}, \ref{fig10} and \ref{fig10c}). (3) We predict the potential blind-survey detection amounts (Table \ref{tab6}) for H21As with various limitations. At most optimistic conditions, FAST covers 80 H21As, ASKAP covers 500, and SKA1-Mid covers 600. Although our results are predicted H21A amount, it is convenient to turn them into DLA amount by dividing the proportional constant $\eta$=0.62.

The lack of BRS and DLA samples with "full" information is one of the crucial obstructions to our study. Nevertheless, with more potential H21As discovered in the future, it is a good chance to research H21As' physical essences and environments, as well as to revise the BRSs from a new aspect. And all this progress can propel our knowledge of the Universe, and help us to comprehend its expansion history and the underlying drivers.

\section*{Acknowledgements}
We thank the anonymous referee for the kind comments that help us greatly to improve this paper. We acknowledge support from National SKA Program of China (2022SKA0110202)
and National Natural Science Foundation of China (grants No.
61802428, 11929301).

\section*{Data Availability}
The data we used in this paper are all open on the Internet. Three radio datasets \citep{rigby11mn,gendre10mn} and three DLA datasets \citep{rao06apj,turns15mn,rao17mn} are available in VizieR. The rest DLAs from \cite{neele16apj,kanek14mn,curran16mn,curran21mn} are directly extracted from their articles.



\bibliographystyle{mnras}
\bibliography{lczp3} 

\begin{thebibliography}{}
\makeatletter
\relax
\def\mn@urlcharsother{\let\do\@makeother \do\$\do\&\do\#\do\^\do\_\do\%\do\~}
\def\mn@doi{\begingroup\mn@urlcharsother \@ifnextchar [ {\mn@doi@}
  {\mn@doi@[]}}
\def\mn@doi@[#1]#2{\def\@tempa{#1}\ifx\@tempa\@empty \href
  {http://dx.doi.org/#2} {doi:#2}\else \href {http://dx.doi.org/#2} {#1}\fi
  \endgroup}
\def\mn@eprint#1#2{\mn@eprint@#1:#2::\@nil}
\def\mn@eprint@arXiv#1{\href {http://arxiv.org/abs/#1} {{\tt arXiv:#1}}}
\def\mn@eprint@dblp#1{\href {http://dblp.uni-trier.de/rec/bibtex/#1.xml}
  {dblp:#1}}
\def\mn@eprint@#1:#2:#3:#4\@nil{\def\@tempa {#1}\def\@tempb {#2}\def\@tempc
  {#3}\ifx \@tempc \@empty \let \@tempc \@tempb \let \@tempb \@tempa \fi \ifx
  \@tempb \@empty \def\@tempb {arXiv}\fi \@ifundefined
  {mn@eprint@\@tempb}{\@tempb:\@tempc}{\expandafter \expandafter \csname
  mn@eprint@\@tempb\endcsname \expandafter{\@tempc}}}

\bibitem[\protect\citeauthoryear{{Abazajian} et~al.,}{{Abazajian}
  et~al.}{2009}]{abaza09apjs}
{Abazajian} K.~N.,  et~al., 2009, \mn@doi [\apjs]
  {10.1088/0067-0049/182/2/543}, \href
  {https://ui.adsabs.harvard.edu/abs/2009ApJS..182..543A} {182, 543}

\bibitem[\protect\citeauthoryear{{Alam} et~al.,}{{Alam}
  et~al.}{2015}]{alam15apjs}
{Alam} S.,  et~al., 2015, \mn@doi [\apjs] {10.1088/0067-0049/219/1/12}, \href
  {https://ui.adsabs.harvard.edu/abs/2015ApJS..219...12A} {219, 12}

\bibitem[\protect\citeauthoryear{{Allison}}{{Allison}}{2021}]{allison21mn}
{Allison} J.~R.,  2021, \mn@doi [\mnras] {10.1093/mnras/stab518}, \href
  {https://ui.adsabs.harvard.edu/abs/2021MNRAS.503..985A} {503, 985}

\bibitem[\protect\citeauthoryear{{Allison} et~al.,}{{Allison}
  et~al.}{2020}]{allison20mn}
{Allison} J.~R.,  et~al., 2020, \mn@doi [\mnras] {10.1093/mnras/staa949}, \href
  {https://ui.adsabs.harvard.edu/abs/2020MNRAS.494.3627A} {494, 3627}

\bibitem[\protect\citeauthoryear{{Allison} et~al.,}{{Allison}
  et~al.}{2022}]{allison22pasa}
{Allison} J.~R.,  et~al., 2022, \mn@doi [\pasa] {10.1017/pasa.2022.3}, \href
  {https://ui.adsabs.harvard.edu/abs/2022PASA...39...10A} {39, e010}

\bibitem[\protect\citeauthoryear{{Bird}, {Garnett}  \& {Ho}}{{Bird}
  et~al.}{2017}]{bird17mn}
{Bird} S.,  {Garnett} R.,   {Ho} S.,  2017, \mn@doi [\mnras]
  {10.1093/mnras/stw3246}, \href
  {https://ui.adsabs.harvard.edu/abs/2017MNRAS.466.2111B} {466, 2111}

\bibitem[\protect\citeauthoryear{{Bolejko}, {Wang}  \& {Lewis}}{{Bolejko}
  et~al.}{2019}]{bolej19arx}
{Bolejko} K.,  {Wang} C.,   {Lewis} G.~F.,  2019, arXiv e-prints, \href
  {https://ui.adsabs.harvard.edu/abs/2019arXiv190704495B} {p. arXiv:1907.04495}

\bibitem[\protect\citeauthoryear{{Braun}}{{Braun}}{2012}]{braun12apj}
{Braun} R.,  2012, \mn@doi [\apj] {10.1088/0004-637X/749/1/87}, \href
  {https://ui.adsabs.harvard.edu/abs/2012ApJ...749...87B} {749, 87}

\bibitem[\protect\citeauthoryear{{Brookes}, {Best}, {Peacock}, {R{\"o}ttgering}
   \& {Dunlop}}{{Brookes} et~al.}{2008}]{brook08mn}
{Brookes} M.~H.,  {Best} P.~N.,  {Peacock} J.~A.,  {R{\"o}ttgering} H.~J.~A.,
  {Dunlop} J.~S.,  2008, \mn@doi [\mnras] {10.1111/j.1365-2966.2008.12786.x},
  \href {https://ui.adsabs.harvard.edu/abs/2008MNRAS.385.1297B} {385, 1297}

\bibitem[\protect\citeauthoryear{{Buchert}, {van Elst}  \&
  {Heinesen}}{{Buchert} et~al.}{2022}]{thoma22arx}
{Buchert} T.,  {van Elst} H.,   {Heinesen} A.,  2022, arXiv e-prints, \href
  {https://ui.adsabs.harvard.edu/abs/2022arXiv220210798B} {p. arXiv:2202.10798}

\bibitem[\protect\citeauthoryear{{Chakrabarti} et~al.,}{{Chakrabarti}
  et~al.}{2022}]{chakr22arx}
{Chakrabarti} S.,  et~al., 2022, arXiv e-prints, \href
  {https://ui.adsabs.harvard.edu/abs/2022arXiv220305924C} {p. arXiv:2203.05924}

\bibitem[\protect\citeauthoryear{{Codur} \& {Marinoni}}{{Codur} \&
  {Marinoni}}{2021}]{codur21prd}
{Codur} R.,  {Marinoni} C.,  2021, \mn@doi [\prd]
  {10.1103/PhysRevD.104.123531}, \href
  {https://ui.adsabs.harvard.edu/abs/2021PhRvD.104l3531C} {104, 123531}

\bibitem[\protect\citeauthoryear{{Condon}, {Cotton}, {Greisen}, {Yin},
  {Perley}, {Taylor}  \& {Broderick}}{{Condon} et~al.}{1998}]{condon98aj}
{Condon} J.~J.,  {Cotton} W.~D.,  {Greisen} E.~W.,  {Yin} Q.~F.,  {Perley}
  R.~A.,  {Taylor} G.~B.,   {Broderick} J.~J.,  1998, \mn@doi [\aj]
  {10.1086/300337}, \href
  {https://ui.adsabs.harvard.edu/abs/1998AJ....115.1693C} {115, 1693}

\bibitem[\protect\citeauthoryear{{Cooke}}{{Cooke}}{2020}]{cooke20mn}
{Cooke} R.,  2020, \mn@doi [\mnras] {10.1093/mnras/stz3465}, \href
  {https://ui.adsabs.harvard.edu/abs/2020MNRAS.492.2044C} {492, 2044}

\bibitem[\protect\citeauthoryear{{Curran}}{{Curran}}{2017a}]{curran17mn}
{Curran} S.~J.,  2017a, \mn@doi [\mnras] {10.1093/mnras/stx933}, \href
  {https://ui.adsabs.harvard.edu/abs/2017MNRAS.470.3159C} {470, 3159}

\bibitem[\protect\citeauthoryear{{Curran}}{{Curran}}{2017b}]{curran17aa}
{Curran} S.~J.,  2017b, \mn@doi [\aap] {10.1051/0004-6361/201731666}, \href
  {https://ui.adsabs.harvard.edu/abs/2017A&A...606A..56C} {606, A56}

\bibitem[\protect\citeauthoryear{{Curran}}{{Curran}}{2021}]{curran21mn}
{Curran} S.~J.,  2021, \mn@doi [\mnras] {10.1093/mnras/stab1865}, \href
  {https://ui.adsabs.harvard.edu/abs/2021MNRAS.506.1548C} {506, 1548}

\bibitem[\protect\citeauthoryear{{Curran}, {Duchesne}, {Divoli}  \&
  {Allison}}{{Curran} et~al.}{2016}]{curran16mn}
{Curran} S.~J.,  {Duchesne} S.~W.,  {Divoli} A.,   {Allison} J.~R.,  2016,
  \mn@doi [\mnras] {10.1093/mnras/stw1938}, \href
  {https://ui.adsabs.harvard.edu/abs/2016MNRAS.462.4197C} {462, 4197}

\bibitem[\protect\citeauthoryear{{Darling}}{{Darling}}{2012}]{darli12apj}
{Darling} J.,  2012, \mn@doi [\apj] {10.1088/2041-8205/761/2/L26}, \href
  {https://ui.adsabs.harvard.edu/abs/2012ApJ...761L..26D} {761, L26}

\bibitem[\protect\citeauthoryear{{Davis} \& {Lineweaver}}{{Davis} \&
  {Lineweaver}}{2001}]{davis01aipc}
{Davis} T.~M.,  {Lineweaver} C.~H.,  2001, in {Durrer} R.,  {Garcia-Bellido}
  J.,   {Shaposhnikov} M.,  eds,  American Institute of Physics Conference
  Series Vol. 555, Cosmology and Particle Physics. pp 348--351 (\mn@eprint
  {arXiv} {astro-ph/0011070}), \mn@doi{10.1063/1.1363540}

\bibitem[\protect\citeauthoryear{{Dong}, {Gonzalez}, {Eikenberry}, {Jeram},
  {Likamonsavad}, {Liske}, {Stelter}  \& {Townsend}}{{Dong}
  et~al.}{2022}]{dong22arx}
{Dong} C.,  {Gonzalez} A.,  {Eikenberry} S.,  {Jeram} S.,  {Likamonsavad} M.,
  {Liske} J.,  {Stelter} D.,   {Townsend} A.,  2022, \mn@doi [\mnras]
  {10.1093/mnras/stac1702}, \href
  {https://ui.adsabs.harvard.edu/abs/2022MNRAS.514.5493D} {514, 5493}

\bibitem[\protect\citeauthoryear{{Dutta}}{{Dutta}}{2019}]{dutta19jaa}
{Dutta} R.,  2019, \mn@doi [Journal of Astrophysics and Astronomy]
  {10.1007/s12036-019-9610-5}, \href
  {https://ui.adsabs.harvard.edu/abs/2019JApA...40...41D} {40, 41}

\bibitem[\protect\citeauthoryear{{Dutta}, {Srianand}, {Gupta}, {Momjian},
  {Noterdaeme}, {Petitjean}  \& {Rahmani}}{{Dutta} et~al.}{2017}]{dutta17mn}
{Dutta} R.,  {Srianand} R.,  {Gupta} N.,  {Momjian} E.,  {Noterdaeme} P.,
  {Petitjean} P.,   {Rahmani} H.,  2017, \mn@doi [\mnras]
  {10.1093/mnras/stw2689}, \href
  {https://ui.adsabs.harvard.edu/abs/2017MNRAS.465..588D} {465, 588}

\bibitem[\protect\citeauthoryear{{Eikenberry} et~al.,}{{Eikenberry}
  et~al.}{2019}]{eiken19baas}
{Eikenberry} S.,  et~al., 2019, in Bulletin of the American Astronomical
  Society. p.~137 (\mn@eprint {arXiv} {1907.08271})

\bibitem[\protect\citeauthoryear{{Esteves}, {Martins}, {Pereira}  \&
  {Alves}}{{Esteves} et~al.}{2021}]{estev21mn}
{Esteves} J.,  {Martins} C.~J.~A.~P.,  {Pereira} B.~G.,   {Alves} C.~S.,  2021,
  \mn@doi [\mnras] {10.1093/mnrasl/slab102}, \href
  {https://ui.adsabs.harvard.edu/abs/2021MNRAS.508L..53E} {508, L53}

\bibitem[\protect\citeauthoryear{{Gaia Collaboration}}{{Gaia
  Collaboration}}{2021}]{gaia21aap}
{Gaia Collaboration} 2021, \mn@doi [\aap] {10.1051/0004-6361/202039734}, \href
  {https://ui.adsabs.harvard.edu/abs/2021A&A...649A...9G} {649, A9}

\bibitem[\protect\citeauthoryear{{Garon} et~al.,}{{Garon}
  et~al.}{2019}]{garon19aj}
{Garon} A.~F.,  et~al., 2019, \mn@doi [\aj] {10.3847/1538-3881/aaff62}, \href
  {https://ui.adsabs.harvard.edu/abs/2019AJ....157..126G} {157, 126}

\bibitem[\protect\citeauthoryear{{Gendre} \& {Wall}}{{Gendre} \&
  {Wall}}{2009}]{gendre09mn}
{Gendre} M.~A.,  {Wall} J.~V.,  2009, \mn@doi [\mnras]
  {10.1111/j.1365-2966.2009.14535.x}, \href
  {https://ui.adsabs.harvard.edu/abs/2009MNRAS.394.1712G} {394, 1712}

\bibitem[\protect\citeauthoryear{{Gendre}, {Best}  \& {Wall}}{{Gendre}
  et~al.}{2010}]{gendre10mn}
{Gendre} M.~A.,  {Best} P.~N.,   {Wall} J.~V.,  2010, \mn@doi [\mnras]
  {10.1111/j.1365-2966.2010.16413.x}, \href
  {https://ui.adsabs.harvard.edu/abs/2010MNRAS.404.1719G} {404, 1719}

\bibitem[\protect\citeauthoryear{{Ger{\'e}b}, {Maccagni}, {Morganti}  \&
  {Oosterloo}}{{Ger{\'e}b} et~al.}{2015}]{gereb15aa}
{Ger{\'e}b} K.,  {Maccagni} F.~M.,  {Morganti} R.,   {Oosterloo} T.~A.,  2015,
  \mn@doi [\aap] {10.1051/0004-6361/201424655}, \href
  {https://ui.adsabs.harvard.edu/abs/2015A&A...575A..44G} {575, A44}

\bibitem[\protect\citeauthoryear{{Grasha}, {Darling}, {Leroy}  \&
  {Bolatto}}{{Grasha} et~al.}{2020}]{grasha20mn}
{Grasha} K.,  {Darling} J.,  {Leroy} A.~K.,   {Bolatto} A.~D.,  2020, \mn@doi
  [\mnras] {10.1093/mnras/staa2521}, \href
  {https://ui.adsabs.harvard.edu/abs/2020MNRAS.498..883G} {498, 883}

\bibitem[\protect\citeauthoryear{{Gupta} et~al.,}{{Gupta}
  et~al.}{2016}]{gupta16mks}
{Gupta} N.,  et~al., 2016, in MeerKAT Science: On the Pathway to the SKA. p.~14
  (\mn@eprint {arXiv} {1708.07371})

\bibitem[\protect\citeauthoryear{{Gupta} et~al.,}{{Gupta}
  et~al.}{2021}]{gupta21apj}
{Gupta} N.,  et~al., 2021, \mn@doi [\apj] {10.3847/1538-4357/abcb85}, \href
  {https://ui.adsabs.harvard.edu/abs/2021ApJ...907...11G} {907, 11}

\bibitem[\protect\citeauthoryear{{Heinesen} \& {Macpherson}}{{Heinesen} \&
  {Macpherson}}{2022}]{heine22jcap}
{Heinesen} A.,  {Macpherson} H.~J.,  2022, \mn@doi [\jcap]
  {10.1088/1475-7516/2022/03/057}, \href
  {https://ui.adsabs.harvard.edu/abs/2022JCAP...03..057H} {2022, 057}

\bibitem[\protect\citeauthoryear{{Jiao}, {Zhang}, {Zhang}, {Yu}, {Zhu}  \&
  {Li}}{{Jiao} et~al.}{2020}]{jiao20jcap}
{Jiao} K.,  {Zhang} J.-C.,  {Zhang} T.-J.,  {Yu} H.-R.,  {Zhu} M.,   {Li} D.,
  2020, \mn@doi [\jcap] {10.1088/1475-7516/2020/01/054}, \href
  {https://ui.adsabs.harvard.edu/abs/2020JCAP...01..054J} {2020, 054}

\bibitem[\protect\citeauthoryear{{Kanekar} et~al.,}{{Kanekar}
  et~al.}{2014}]{kanek14mn}
{Kanekar} N.,  et~al., 2014, \mn@doi [\mnras] {10.1093/mnras/stt2338}, \href
  {https://ui.adsabs.harvard.edu/abs/2014MNRAS.438.2131K} {438, 2131}

\bibitem[\protect\citeauthoryear{{Kloeckner} et~al.,}{{Kloeckner}
  et~al.}{2015}]{klo15aaska}
{Kloeckner} H.~R.,  et~al., 2015, in Advancing Astrophysics with the Square
  Kilometre Array (AASKA14). p.~27 (\mn@eprint {arXiv} {1501.03822})

\bibitem[\protect\citeauthoryear{{Koribalski} et~al.,}{{Koribalski}
  et~al.}{2020}]{kori20apss}
{Koribalski} B.~S.,  et~al., 2020, \mn@doi [\apss]
  {10.1007/s10509-020-03831-4}, \href
  {https://ui.adsabs.harvard.edu/abs/2020Ap&SS.365..118K} {365, 118}

\bibitem[\protect\citeauthoryear{{Li} et~al.,}{{Li} et~al.}{2018}]{li18imm}
{Li} D.,  et~al., 2018, \mn@doi [IEEE Microwave Magazine]
  {10.1109/MMM.2018.2802178}, \href
  {https://ui.adsabs.harvard.edu/abs/2018IMMag..19..112L} {19, 112}

\bibitem[\protect\citeauthoryear{{Liske} et~al.,}{{Liske}
  et~al.}{2008}]{liske08mn}
{Liske} J.,  et~al., 2008, \mn@doi [\mnras] {10.1111/j.1365-2966.2008.13090.x},
  \href {https://ui.adsabs.harvard.edu/abs/2008MNRAS.386.1192L} {386, 1192}

\bibitem[\protect\citeauthoryear{{Loeb}}{{Loeb}}{1998}]{loeb98apj}
{Loeb} A.,  1998, \mn@doi [\apj] {10.1086/311375}, \href
  {https://ui.adsabs.harvard.edu/abs/1998ApJ...499L.111L} {499, L111}

\bibitem[\protect\citeauthoryear{{Lu}, {Jiao}, {Zhang}, {Zhang}  \& {Zhu}}{{Lu}
  et~al.}{2022}]{lu21arx}
{Lu} C.-Z.,  {Jiao} K.,  {Zhang} T.,  {Zhang} T.-J.,   {Zhu} M.,  2022, \mn@doi
  [Physics of the Dark Universe] {10.1016/j.dark.2022.101088}, \href
  {https://ui.adsabs.harvard.edu/abs/2022PDU....3701088L} {37, 101088}

\bibitem[\protect\citeauthoryear{{Ma} et~al.,}{{Ma} et~al.}{2019}]{ma19apjs}
{Ma} Z.,  et~al., 2019, \mn@doi [\apjs] {10.3847/1538-4365/aaf9a2}, \href
  {https://ui.adsabs.harvard.edu/abs/2019ApJS..240...34M} {240, 34}

\bibitem[\protect\citeauthoryear{{Marcos-Caballero}, {Vielva},
  {Martinez-Gonzalez}, {Finelli}, {Gruppuso}  \& {Schiavon}}{{Marcos-Caballero}
  et~al.}{2013}]{marcab13arx}
{Marcos-Caballero} A.,  {Vielva} P.,  {Martinez-Gonzalez} E.,  {Finelli} F.,
  {Gruppuso} A.,   {Schiavon} F.,  2013, arXiv e-prints, \href
  {https://ui.adsabs.harvard.edu/abs/2013arXiv1312.0530M} {p. arXiv:1312.0530}

\bibitem[\protect\citeauthoryear{{Matthews}, {Condon}, {Cotton}  \&
  {Mauch}}{{Matthews} et~al.}{2021}]{matth21apj}
{Matthews} A.~M.,  {Condon} J.~J.,  {Cotton} W.~D.,   {Mauch} T.,  2021,
  \mn@doi [\apj] {10.3847/1538-4357/abdd37}, \href
  {https://ui.adsabs.harvard.edu/abs/2021ApJ...909..193M} {909, 193}

\bibitem[\protect\citeauthoryear{{Melia}}{{Melia}}{2022}]{melia22ejph}
{Melia} F.,  2022, \mn@doi [European Journal of Physics]
  {10.1088/1361-6404/ac4646}, \href
  {https://ui.adsabs.harvard.edu/abs/2022EJPh...43c5601M} {43, 035601}

\bibitem[\protect\citeauthoryear{{Mishra}}{{Mishra}}{2022}]{mishr22prd}
{Mishra} P.,  2022, \mn@doi [\prd] {10.1103/PhysRevD.105.063520}, \href
  {https://ui.adsabs.harvard.edu/abs/2022PhRvD.105f3520M} {105, 063520}

\bibitem[\protect\citeauthoryear{{Moresco} et~al.,}{{Moresco}
  et~al.}{2022}]{mores22arx}
{Moresco} M.,  et~al., 2022, arXiv e-prints, \href
  {https://ui.adsabs.harvard.edu/abs/2022arXiv220107241M} {p. arXiv:2201.07241}

\bibitem[\protect\citeauthoryear{{Murray}, {Stanimirovi{\'c}}, {Goss},
  {Heiles}, {Dickey}, {Babler}  \& {Kim}}{{Murray} et~al.}{2018}]{murray18apjs}
{Murray} C.~E.,  {Stanimirovi{\'c}} S.,  {Goss} W.~M.,  {Heiles} C.,  {Dickey}
  J.~M.,  {Babler} B.,   {Kim} C.-G.,  2018, \mn@doi [\apjs]
  {10.3847/1538-4365/aad81a}, \href
  {https://ui.adsabs.harvard.edu/abs/2018ApJS..238...14M} {238, 14}

\bibitem[\protect\citeauthoryear{{Neeleman}, {Prochaska}, {Ribaudo}, {Lehner},
  {Howk}, {Rafelski}  \& {Kanekar}}{{Neeleman} et~al.}{2016}]{neele16apj}
{Neeleman} M.,  {Prochaska} J.~X.,  {Ribaudo} J.,  {Lehner} N.,  {Howk} J.~C.,
  {Rafelski} M.,   {Kanekar} N.,  2016, \mn@doi [\apj]
  {10.3847/0004-637X/818/2/113}, \href
  {https://ui.adsabs.harvard.edu/abs/2016ApJ...818..113N} {818, 113}

\bibitem[\protect\citeauthoryear{{Noterdaeme} et~al.,}{{Noterdaeme}
  et~al.}{2012}]{noter12aa}
{Noterdaeme} P.,  et~al., 2012, \mn@doi [\aap] {10.1051/0004-6361/201220259},
  \href {https://ui.adsabs.harvard.edu/abs/2012A&A...547L...1N} {547, L1}

\bibitem[\protect\citeauthoryear{{Planck Collaboration}}{{Planck
  Collaboration}}{2020}]{planck20aa}
{Planck Collaboration} 2020, \mn@doi [\aap] {10.1051/0004-6361/201833910},
  \href {https://ui.adsabs.harvard.edu/abs/2020A&A...641A...6P} {641, A6}

\bibitem[\protect\citeauthoryear{{Prochaska} \& {Wolfe}}{{Prochaska} \&
  {Wolfe}}{2009}]{procha09apj}
{Prochaska} J.~X.,  {Wolfe} A.~M.,  2009, \mn@doi [\apj]
  {10.1088/0004-637X/696/2/1543}, \href
  {https://ui.adsabs.harvard.edu/abs/2009ApJ...696.1543P} {696, 1543}

\bibitem[\protect\citeauthoryear{{Rao}, {Turnshek}  \& {Nestor}}{{Rao}
  et~al.}{2006}]{rao06apj}
{Rao} S.~M.,  {Turnshek} D.~A.,   {Nestor} D.~B.,  2006, \mn@doi [\apj]
  {10.1086/498132}, \href
  {https://ui.adsabs.harvard.edu/abs/2006ApJ...636..610R} {636, 610}

\bibitem[\protect\citeauthoryear{{Rao}, {Turnshek}, {Sardane}  \&
  {Monier}}{{Rao} et~al.}{2017}]{rao17mn}
{Rao} S.~M.,  {Turnshek} D.~A.,  {Sardane} G.~M.,   {Monier} E.~M.,  2017,
  \mn@doi [\mnras] {10.1093/mnras/stx1787}, \href
  {https://ui.adsabs.harvard.edu/abs/2017MNRAS.471.3428R} {471, 3428}

\bibitem[\protect\citeauthoryear{{Rigby}, {Best}, {Brookes}, {Peacock},
  {Dunlop}, {R{\"o}ttgering}, {Wall}  \& {Ker}}{{Rigby}
  et~al.}{2011}]{rigby11mn}
{Rigby} E.~E.,  {Best} P.~N.,  {Brookes} M.~H.,  {Peacock} J.~A.,  {Dunlop}
  J.~S.,  {R{\"o}ttgering} H.~J.~A.,  {Wall} J.~V.,   {Ker} L.,  2011, \mn@doi
  [\mnras] {10.1111/j.1365-2966.2011.19167.x}, \href
  {https://ui.adsabs.harvard.edu/abs/2011MNRAS.416.1900R} {416, 1900}

\bibitem[\protect\citeauthoryear{{Sadler} et~al.,}{{Sadler}
  et~al.}{2020}]{sadler20mn}
{Sadler} E.~M.,  et~al., 2020, \mn@doi [\mnras] {10.1093/mnras/staa2390}, \href
  {https://ui.adsabs.harvard.edu/abs/2020MNRAS.499.4293S} {499, 4293}

\bibitem[\protect\citeauthoryear{{Sandage}}{{Sandage}}{1962}]{sanda62apj}
{Sandage} A.,  1962, \mn@doi [\apj] {10.1086/147385}, \href
  {https://ui.adsabs.harvard.edu/abs/1962ApJ...136..319S} {136, 319}

\bibitem[\protect\citeauthoryear{{Schmidt}, {Connolly}  \& {Hopkins}}{{Schmidt}
  et~al.}{2006}]{schmi06apj}
{Schmidt} S.~J.,  {Connolly} A.~J.,   {Hopkins} A.~M.,  2006, \mn@doi [\apj]
  {10.1086/506444}, \href
  {https://ui.adsabs.harvard.edu/abs/2006ApJ...649...63S} {649, 63}

\bibitem[\protect\citeauthoryear{{Titov} \& {Kr{\'a}sn{\'a}}}{{Titov} \&
  {Kr{\'a}sn{\'a}}}{2018}]{titov18aap}
{Titov} O.,  {Kr{\'a}sn{\'a}} H.,  2018, \mn@doi [\aap]
  {10.1051/0004-6361/201731901}, \href
  {https://ui.adsabs.harvard.edu/abs/2018A&A...610A..36T} {610, A36}

\bibitem[\protect\citeauthoryear{{Titov}, {Lambert}  \& {Gontier}}{{Titov}
  et~al.}{2011}]{titov11aa}
{Titov} O.,  {Lambert} S.~B.,   {Gontier} A.~M.,  2011, \mn@doi [\aap]
  {10.1051/0004-6361/201015718}, \href
  {https://ui.adsabs.harvard.edu/abs/2011A&A...529A..91T} {529, A91}

\bibitem[\protect\citeauthoryear{{Titov} et~al.,}{{Titov}
  et~al.}{2022}]{titov22mn}
{Titov} O.,  et~al., 2022, \mn@doi [\mnras] {10.1093/mnras/stac038}, \href
  {https://ui.adsabs.harvard.edu/abs/2022MNRAS.512..874T} {512, 874}

\bibitem[\protect\citeauthoryear{{Toba} et~al.,}{{Toba}
  et~al.}{2019}]{toba19apjs}
{Toba} Y.,  et~al., 2019, \mn@doi [\apjs] {10.3847/1538-4365/ab238d}, \href
  {https://ui.adsabs.harvard.edu/abs/2019ApJS..243...15T} {243, 15}

\bibitem[\protect\citeauthoryear{{Turnshek}, {Monier}, {Rao}, {Hamilton},
  {Sardane}  \& {Held}}{{Turnshek} et~al.}{2015}]{turns15mn}
{Turnshek} D.~A.,  {Monier} E.~M.,  {Rao} S.~M.,  {Hamilton} T.~S.,  {Sardane}
  G.~M.,   {Held} R.,  2015, \mn@doi [\mnras] {10.1093/mnras/stv224}, \href
  {https://ui.adsabs.harvard.edu/abs/2015MNRAS.449.1536T} {449, 1536}

\bibitem[\protect\citeauthoryear{{Waddington}, {Dunlop}, {Peacock}  \&
  {Windhorst}}{{Waddington} et~al.}{2001}]{waddi01mn}
{Waddington} I.,  {Dunlop} J.~S.,  {Peacock} J.~A.,   {Windhorst} R.~A.,  2001,
  \mn@doi [\mnras] {10.1046/j.1365-8711.2001.04934.x}, \href
  {https://ui.adsabs.harvard.edu/abs/2001MNRAS.328..882W} {328, 882}

\bibitem[\protect\citeauthoryear{{Weltman} et~al.,}{{Weltman}
  et~al.}{2020}]{welt20pasa}
{Weltman} A.,  et~al., 2020, \mn@doi [\pasa] {10.1017/pasa.2019.42}, \href
  {https://ui.adsabs.harvard.edu/abs/2020PASA...37....2W} {37, e002}

\bibitem[\protect\citeauthoryear{{White}, {Becker}, {Helfand}  \&
  {Gregg}}{{White} et~al.}{1997}]{white97apj}
{White} R.~L.,  {Becker} R.~H.,  {Helfand} D.~J.,   {Gregg} M.~D.,  1997,
  \mn@doi [\apj] {10.1086/303564}, \href
  {https://ui.adsabs.harvard.edu/abs/1997ApJ...475..479W} {475, 479}

\bibitem[\protect\citeauthoryear{{Wright} \& {Eastman}}{{Wright} \&
  {Eastman}}{2014}]{wirght14pasp}
{Wright} J.~T.,  {Eastman} J.~D.,  2014, \mn@doi [\pasp] {10.1086/678541},
  \href {https://ui.adsabs.harvard.edu/abs/2014PASP..126..838W} {126, 838}

\bibitem[\protect\citeauthoryear{{Wu}, {Haynes}, {Giovanelli}, {Zhu}  \&
  {Chen}}{{Wu} et~al.}{2015}]{wu15aas}
{Wu} Z.~Z.,  {Haynes} M.~P.,  {Giovanelli} R.,  {Zhu} M.,   {Chen} R.~R.,
  2015, Acta Astronomica Sinica, \href
  {https://ui.adsabs.harvard.edu/abs/2015AcASn..56..112W} {56, 112}

\bibitem[\protect\citeauthoryear{{Xu} \& {Han}}{{Xu} \& {Han}}{2014}]{xu14mn}
{Xu} J.,  {Han} J.~L.,  2014, \mn@doi [\mnras] {10.1093/mnras/stu1018}, \href
  {https://ui.adsabs.harvard.edu/abs/2014MNRAS.442.3329X} {442, 3329}

\bibitem[\protect\citeauthoryear{{Xu}, {Wang}  \& {Zhao}}{{Xu}
  et~al.}{2012}]{xu12aa}
{Xu} M.~H.,  {Wang} G.~L.,   {Zhao} M.,  2012, \mn@doi [\aap]
  {10.1051/0004-6361/201219593}, \href
  {https://ui.adsabs.harvard.edu/abs/2012A&A...544A.135X} {544, A135}

\bibitem[\protect\citeauthoryear{{Yu}, {Zhang}  \& {Pen}}{{Yu}
  et~al.}{2014}]{yu14prl}
{Yu} H.-R.,  {Zhang} T.-J.,   {Pen} U.-L.,  2014, \mn@doi [\prl]
  {10.1103/PhysRevLett.113.041303}, \href
  {https://ui.adsabs.harvard.edu/abs/2014PhRvL.113d1303Y} {113, 041303}

\bibitem[\protect\citeauthoryear{{Yu}, {Pen}, {Zhang}, {Li}  \& {Chen}}{{Yu}
  et~al.}{2017}]{yu17raa}
{Yu} H.-R.,  {Pen} U.-L.,  {Zhang} T.-J.,  {Li} D.,   {Chen} X.,  2017, \mn@doi
  [\raa] {10.1088/1674-4527/17/6/49}, \href
  {https://ui.adsabs.harvard.edu/abs/2017RAA....17...49Y} {17, 049}

\bibitem[\protect\citeauthoryear{{Zakamska} \& {Tremaine}}{{Zakamska} \&
  {Tremaine}}{2005}]{zakam05aj}
{Zakamska} N.~L.,  {Tremaine} S.,  2005, \mn@doi [\aj] {10.1086/444476}, \href
  {https://ui.adsabs.harvard.edu/abs/2005AJ....130.1939Z} {130, 1939}

\bibitem[\protect\citeauthoryear{{Zhang} et~al.,}{{Zhang}
  et~al.}{2019}]{zhang19scpma}
{Zhang} K.,  et~al., 2019, \mn@doi [Science China Physics, Mechanics, and
  Astronomy] {10.1007/s11433-019-9383-y}, \href
  {https://ui.adsabs.harvard.edu/abs/2019SCPMA..6259506Z} {62, 959506}

\bibitem[\protect\citeauthoryear{{Zhang}, {Zhu}, {Wu}, {Yu}, {Jiang}, {Yue},
  {Huang}  \& {Hao}}{{Zhang} et~al.}{2021}]{zhang21mn}
{Zhang} B.,  {Zhu} M.,  {Wu} Z.-Z.,  {Yu} Q.-Z.,  {Jiang} P.,  {Yue} Y.-L.,
  {Huang} M.-L.,   {Hao} Q.-L.,  2021, \mn@doi [\mnras]
  {10.1093/mnras/stab754}, \href
  {https://ui.adsabs.harvard.edu/abs/2021MNRAS.503.5385Z} {503, 5385}

\bibitem[\protect\citeauthoryear{{Zwaan}, {van der Hulst}, {Briggs},
  {Verheijen}  \& {Ryan-Weber}}{{Zwaan} et~al.}{2005}]{zwaan05mn}
{Zwaan} M.~A.,  {van der Hulst} J.~M.,  {Briggs} F.~H.,  {Verheijen} M.~A.~W.,
   {Ryan-Weber} E.~V.,  2005, \mn@doi [\mnras]
  {10.1111/j.1365-2966.2005.09698.x}, \href
  {https://ui.adsabs.harvard.edu/abs/2005MNRAS.364.1467Z} {364, 1467}

\bibitem[\protect\citeauthoryear{{de Zotti}, {Massardi}, {Negrello}  \&
  {Wall}}{{de Zotti} et~al.}{2010}]{dezot10aapr}
{de Zotti} G.,  {Massardi} M.,  {Negrello} M.,   {Wall} J.,  2010, \mn@doi
  [\aapr] {10.1007/s00159-009-0026-0}, \href
  {https://ui.adsabs.harvard.edu/abs/2010A&ARv..18....1D} {18, 1}

\makeatother
\end{thebibliography}



\bsp	
\label{lastpage}
\end{document}